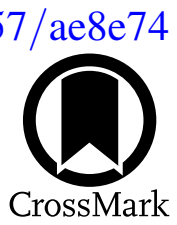

# Time-resolved JWST Retrieval Analysis of the Coldest Brown Dwarf: Temperature and Chemical Variations in WISE 0855−07

Harshil Kothari[1], Caroline V. Morley[2], Brittany E. Miles[3], Melanie J. Rowland[4], Natasha Batalha[5], Michael C. Cushing[1], Andrew J. Skemer[5], James Mang[6], Brianna Lacy[7,8], Johanna M. Vos[9], Channon Visscher[10,11], Adam C. Schneider[12], Genaro Suarez[13], Mikayla J. Wilson[14], and Allison M. McCarthy[9]
[1] Ritter Astrophysical Research Center, Department of Physics & Astronomy, University of Toledo, 2801 W. Bancroft St., Toledo, OH 43606, USA
[2] University of Texas at Austin, Department of Astronomy, 2515 Speedway C1400, Austin, TX 78712, USA
[3] Steward Observatory, University of Arizona, 933 N. Cherry Ave., Tucson, AZ 85721, USA
[4] Department of Astrophysics, American Museum of Natural History, New York, NY 10024, USA
[5] Space Science and Astrobiology Division, NASA Ames Research Center, Moffett Field, CA 94035, USA
[6] Department of Astronomy, University of Texas at Austin, Austin, TX 78712, USA
[7] Oak Ridge Associated Universities, USA
[8] NASA Ames Research Center, USA
[9] School of Physics, Trinity College Dublin, The University of Dublin, Dublin 2, Ireland
[10] Chemistry & Planetary Sciences, Dordt University, Sioux Center, IA, USA
[11] Center for Extrasolar Planetary Systems, Space Science Institute, Boulder, CO, USA
[12] United States Naval Observatory, Flagstaff Station, 10391 West Naval Observatory Rd., Flagstaff, AZ 86005, USA
[13] Department of Astrophysics, American Museum of Natural History, Central Park West at 79th Street, New York, NY 10024, USA
[14] Department of Astronomy & Astrophysics, University of California, Santa Cruz, CA 95064, USA

## Abstract

WISE J0855–07 is the coldest known brown dwarf (estimated $T_{\rm eff} \sim 250$ K), offering a rare opportunity to probe atmospheric physics in the temperature and mass regime that overlaps directly with exoplanets. At these temperatures, modest perturbations to the thermal structure and chemical abundances can produce measurable changes in the emergent mid-infrared spectrum, making time-resolved spectroscopy a powerful way to test whether the atmosphere is longitudinally and temporally homogeneous. Using James Webb Space Telescope NIRSpec/G395M time-series spectra spanning ∼3 to 5.2 $\mu$m over 11 hr, we analyze every third spectrum from the dataset with a series of time-resolved atmospheric retrievals using the `Brewster` retrieval framework. Using this approach, we model epoch-to-epoch changes in the atmospheric state that drive the observed variability, including the thermal structure and key molecular abundances. From the time-series retrievals, we find that the primary source of the observed spectral variability is driven by the variations in the thermal structure, which range from ∼1 to <100 K from epoch to epoch, particularly between the pressures of ∼0.4 and 10 bar. In addition, changes in the abundances of CO, and possibly $PH_3$ and $CO_2$, further modulate the spectrum within specific molecular features. These measurements provide a retrieval-based view of atmospheric variability in an ultracold, planet-like object, directly constraining how thermal and chemical structure evolve on observable timescales.

## 1. Introduction

WISE J085510.83−071442.5 (hereafter WISE 0855−07) is the coldest known isolated substellar object. At an effective temperature of only ∼250 K, it provides a unique bridge between the atmospheres of solar system giant planets and those of warmer brown dwarfs and gas giant exoplanets. The object was identified by K. L. Luhman (2014) in Wide-field Infrared Survey Explorer (WISE; E. L. Wright et al. 2010) survey images via its large proper motion of $\mu = 8\farcs1 \pm 0\farcs1$ yr$^{-1}$ and extremely red mid-infrared colors of [3.6]–[4.5] = 3.55 ± 0.05 mag (from [3.6] = 17.44 ± 0.05 mag and [4.5] = 13.89 ± 0.02 mag). Subsequent Spitzer and ground-based follow-up observations established its very low luminosity of $\log(L/L_\odot) = -8.57$ (i.e., $L \simeq 2.7 \times 10^{-9} L_\odot$) and very low effective temperature range of $T_{\rm eff}$= 225−250 K (M. R. Zapatero Osorio et al. 2016).

Early observational work on WISE 0855−07 focused on broadband photometry (J. C. Beamín et al. 2014; J. K. Faherty et al. 2014; T. G. Kopytova et al. 2014; K. L. Luhman & T. L. Esplin 2016; A. C. Schneider et al. 2016), which provided low signal-to-noise ratio (S/N) measurements in the near-infrared and revealed an object that is exceptionally faint at short wavelengths yet comparatively bright in the mid-infrared. This extreme spectral energy distribution reflects its remarkably low effective temperature (∼250 K), at which strong methane and water vapor absorption suppresses emergent flux in the near-infrared. As emphasized by A. C. Schneider et al., WISE 0855−07 lies at the "collapse of the Wien tail," where the Planck function falls off so steeply that very little intrinsic flux emerges at near-IR wavelengths. Subsequent ground-based spectroscopy at 5 $\mu$m by A. J. Skemer et al. provided the first direct spectroscopic window into its atmosphere, probing deeper, warmer layers revealing prominent molecular absorption features consistent with water vapor and methane. At these temperatures, condensate cloud formation is also expected to play a major role; models by A. Burrows et al. (2003), C. V. Morley et al. (2014b), and B. Lacy & A. Burrows (2023) predict the formation of water-ice clouds, which can further shape the emergent spectrum and introduce spatial heterogeneity. Time-series photometric monitoring with Spitzer revealed variability at the few-percent level, pointing to heterogeneous cloud cover or patchy thermal structures (K. L. Luhman & T. L. Esplin 2016).

With the launch of the James Webb Space Telescope (JWST), we now have access to a broad wavelength coverage using the Near Infrared Spectrograph (hereafter NIRSpec; P. Jakobsen et al. 2022) and the Mid-Infrared Instrument (hereafter MIRI; G. H. Rieke et al. 2015) that allows us to observe near- and mid-infrared wavelength regime with a high S/N, enabling the first detailed chemical abundance constraints for the coldest known brown dwarf (K. L. Luhman 2014). Recent JWST datasets and retrieval analyses of cold brown dwarfs have reported detections and constraints for major molecules like $H_2O$, $CH_4$, and $NH_3$ (H. Kothari et al. 2024; B. W. P. Lew et al. 2024; M. J. Rowland et al. 2024; H. Kühnle et al. 2025), isotopologues (e.g., $CH_3D$) (M. J. Rowland et al. 2024), and trace species like $PH_3$ (M. J. Rowland et al. 2024; A. J. Burgasser et al. 2025) and $SiH_4$ (J. K. Faherty et al. 2025). Among the trace species probed in the 3–5 $\mu$m region, $^{12}CO$, $CO_2$, and $PH_3$ are especially valuable because they have a long observational legacy as tracers of disequilibrium chemistry in planetary and substellar atmospheres, from the early detections of CO and $PH_3$ in Jupiter (R. Beer 1975; H. P. Larson et al. 1977) to the detection of CO in the brown dwarf Gliese 229B (K. S. Noll et al. 1997) and the unambiguous detection of $CO_2$ in the exoplanet WASP-39b with JWST (E.-M. Ahrer et al. 2023). These studies demonstrate that JWST spectroscopy can constrain many of the major carbon-, nitrogen-, and oxygen-bearing molecules, and even isotopic ratios, in objects with temperatures of only a few hundred kelvin. However, these studies have largely focused on either single-epoch or time-averaged spectra, while time-resolved spectroscopic analysis of the coldest brown dwarfs remains comparatively unexplored.

### 1.1. Miles et al. (2026, submitted)

Time-resolved observations provide a powerful diagnostic of atmospheric heterogeneity, as variability encodes changes in temperature, composition, and cloud opacity across the surface. In this context, Miles et al. (2026, submitted) presented the first spectroscopic time-series observations of WISE 0855−07 using JWST/NIRSpec with the G395M/F290LP configuration over an 11 hr baseline. They reported wavelength-dependent variability with peak-to-peak amplitudes of 1%–6%. Spectral regions dominated by strong $H_2O$, $CH_4$, and $NH_3$ opacities exhibit suppressed variability, whereas the largest peak-to-peak variability amplitudes occur within CO absorption bands. At wavelengths where $CH_4$, $NH_3$, and $H_2O$ opacities are comparatively weak, variability within disequilibrium species such as $^{12}CO$, $CO_2$, and $PH_3$ absorption bands emerges, suggesting sensitivity to deeper atmospheric temperature perturbations.

Using 1D radiative-convective equilibrium forward model comparisons and $\chi^2$ analysis, Miles et al. found that models including water clouds improve fits to the mean spectrum, with the best-fitting solution corresponding to a cloudy atmosphere with moderately high vertical mixing (represented by $K_{zz}$) ($T_{\rm eff}$ $\sim 250$ K; $\log(K_{zz}) = 6$ cm$^2$ s$^{-1}$). However, persistently high $\chi^2$ values and differences in cloud vertical structure between atmospheric model grids highlight significant degeneracies inherent to forward modeling. Modeling of the observed variability further required coupled changes in effective temperature, water cloud thickness, and the abundances of $^{12}CO$, $CO_2$, and $PH_3$, implying contributions from multiple pressure levels and potentially zonal-jet-dominated atmospheric dynamics.

### 1.2. This Work

We analyze the same JWST/NIRSpec time-series dataset but adopt a Bayesian atmospheric retrieval framework rather than a forward model grid comparison approach. While forward models explore discrete regions of parameter space to generate synthetic spectra for any combination of parameters within the model grid under fixed assumptions about cloud structure, chemistry, and vertical mixing, retrievals allow the volume number mixing ratios (VMRs),[15] thermal structure, and cloud properties to vary more flexibly, without imposing strong a priori constraints on the underlying physics and chemistry. To investigate the mechanisms driving the observed variability in WISE 0855−07, we test seven variability retrieval scenarios.

Relative to previous time-series retrieval studies of other brown dwarfs (E. Nasedkin et al. 2025; F. Wang et al. 2026) our dataset combines (1) high spectral resolving power across the $\sim$3–5.2 $\mu$m window, (2) continuous time-sampling that captures temporal modulation, and (3) the S/N necessary for retrievals on individual spectra rather than only an average spectrum. We perform time-series retrievals on WISE 0855−07 time-resolved spectra to map variations in thermal structure and molecular abundances to constrain the observed variability. In Section 2, we describe the JWST NIRSpec/G395M observations, data reduction, and extraction of the 44 time-series spectra. Section 3 outlines the forward model, the retrieval framework, and the statistical methods used to assess the quality of the time-series retrievals. Section 4 presents the time-series retrieval results using VMR-only, temperature-only, and a combination of both for variability modeling scenarios. In Section 5, we discuss the temporal changes in VMR of CO, $PH_3$, and $CO_2$, and the temperature structure for the part of the atmosphere probed by our time-series dataset. We also compare our mean spectral retrieval with the previous retrieval work done by M. J. Rowland et al. (2024), which uses the mean spectrum from the same dataset for their retrieval analysis. Lastly, we present the main takeaways of this paper in Section 6.

## 2. Time-series Spectra

We use JWST NIRSpec (P. Jakobsen et al. 2022) Bright Object Time Series observations of WISE 0855–07 obtained as a part of Program ID: 2327 (PI: Andrew J. Skemer, Co-PIs: Caroline V. Morley and Brittany E. Miles). The S1600A1 slit ($1''.6 \times 1''.6$) was used with the G395M/F290LP grating/filter combination, providing continuous coverage from 2.87 to 5.10 $\mu$m at a spectral resolving power ($R \equiv \lambda/\Delta\lambda$) ranging from $\sim$700 to 1300. The observations span $\sim$11 hr and consist

[15] The volume number mixing ratio of a species is the number density of that species divided by the total number density of the gas.

of 44 integrations per exposure with NRSRAPID readout, yielding 15 minute cadence spectra.

Miles et al. (2026, submitted) processed the data with JWST pipeline v1.14.0 (CRDS context jwst_1215.pmap) using default Stage 1 and Stage 2 steps, including reference pixel corrections, bias and linearity corrections, ramp fitting, flat-fielding, wavelength calibration, and 2D spectral extraction. A 4 pixel wide aperture was used for extraction, with background estimated from surrounding pixels. Problematic pixels were identified and masked during the reduction. Individual pixels were removed that showed nonphysical behavior relative to neighboring wavelengths across the full time series: 1 pixel at 2.8935 $\mu$m with anomalously high flux (a factor of 2 larger than adjacent points), 1 pixel at 4.1780 $\mu$m with spurious 200% variability, pixels at 4.6920 $\mu$m, 1 pixel at 4.9955 $\mu$m with fluxes systematically higher than surrounding wavelengths, and at 5.1660 $\mu$m with consistently negative fluxes. In addition, Miles et al. isolated 1–2 pixel wide clusters of hot pixels appearing in seven spectra (spectra 17, 28, 29, 30, 32, 37, and 44), which were masked. After all masking, the final product is a time series of 44 flux-calibrated spectra. The wavelength solution remains highly stable, with a median pointing-to-pointing variation of $1.8 \times 10^{-6}$ $\mu$m, which is $\sim$900 times smaller than the median spectral resolution element ($1.7 \times 10^{-3}$ $\mu$m). A more detailed discussion about the observation and data reduction of these spectra can be found in Miles et al. (2026, submitted).

The dominant time-variable feature is associated with $^{12}$CO absorption with a maximum peak-to-peak flux variability of $\sim$8% relative to the mean spectrum. To balance computational efficiency against temporal and model-scenario coverage, we perform retrievals on every third spectrum from the time-series dataset, corresponding to an effective cadence of approximately 45 minutes. This choice preserves sampling over the full $\sim$11 hr sequence while reducing the number of expensive retrievals. An alternative approach would be to coadd each group of three adjacent 15 minute spectra, thereby increasing the S/N by approximately $\sqrt{3}$ in the limit of uncorrelated noise. We instead use individual observed spectra as our fiducial data products because the goal of this analysis is to model time-resolved spectral variability at specific epochs, rather than spectra averaged over finite intervals of rotational phase. Such averaging could dilute intrinsic temporal variability and would not necessarily improve the uncertainties by $\sqrt{3}$ in the presence of correlated calibration or instrumental systematics. Since each flux-calibrated spectrum has a slightly different wavelength solution (differing by up to 0.01 $\mu$m), all spectra were interpolated onto a common wavelength grid defined by the first spectrum in the time-series dataset. We tested linear, cubic, and sinc interpolation schemes by comparing the fluxes of the interpolated spectra against the raw data, and found that cubic interpolation reproduced the observed flux the best.

In Figure 1, the top panel illustrates the wavelength-dependent molecular absorption cross sections at 1 bar and 500 K, the middle panel shows the subset of every third spectrum from the time-series dataset used for retrievals, and the bottom panel corresponds to the observed spectral variability relative to the overall mean spectrum. The middle panel labels the P-, Q-, and R-branches of key molecules that contribute to absorption within the 3.8–5.0 $\mu$m region. For $^{12}$CO, the fundamental ($\nu$ = 1–0) band near 4.5–4.9 $\mu$m exhibits well-defined P- and R-branch features (D. Saumon et al. 2000; R. S. Freedman et al. 2014). The $PH_3$ $\nu_2$ bending mode near 4.1–4.5 $\mu$m partially overlaps the $^{12}$CO P-branch region (C. Sousa-Silva et al. 2015). The $CO_2$ $\nu_3$ antisymmetric stretching mode, centered at 4.26 $\mu$m, produces strong P- and R-branch structure with a weak central Q-branch (X. Huang et al. 2014), which aligns closely with the structured variability seen in the region. Additional broad absorption arises from the $CH_4$ $\nu_3$ fundamental near 3.3 $\mu$m and the blended $H_2O$ lines below 4.2 $\mu$m (K. Lodders & B. Fegley 2006; C. M. Sharp & A. Burrows 2007), though variability there is indistinguishable.

To ensure that the observed variability is statistically significant (i.e., that a peak-to-peak modulation can be distinguished from measurement noise at the required confidence level), we adopt a wavelength cutoff based on the S/N necessary to detect an 8% peak-to-peak variability amplitude. For a per-pixel fractional uncertainty

$$\sigma_f = \frac{1}{\mathrm{S/N}}, \quad (1)$$

a peak-to-peak variability amplitude $A$ is detected at the $n\sigma$ level when

$$n\sigma_f \leqslant A. \quad (2)$$

For a measured peak-to-peak variability amplitude of $A = 0.08$ (8%), this corresponds to a detection threshold of $\mathrm{S/N} \gtrsim 40$ for a $3\sigma$ significance. Since the S/N falls below 40 at wavelengths shorter than 3.8 $\mu$m, we adopt 3.8 $\mu$m as the short-wavelength cutoff. This minimizes the risk of interpreting noise or calibration residuals as astrophysical variability, which is primarily confined between $\sim$3.8 and 5.2 $\mu$m, the wavelength region dominated by $^{12}$CO, $CO_2$, and $PH_3$ molecular features.

As described in Miles et al. (2026, submitted), the observed variability is highly structured and correlated with the absorption bands of major disequilibrium species, especially $^{12}$CO and $CO_2$, as shown by the cross sections in Figure 1 (top panel). The temporal sequence of 15 spectra used for retrieval analysis is indexed in the middle and bottom panels.

## 3. Time-series Retrieval Analysis Methodology

To interpret the time-resolved JWST spectra of WISE 0855−07, we use the `Brewster` atmospheric retrieval framework (B. Burningham et al. 2017), modified here to analyze time-varying spectra. In `Brewster`, a forward model generates emergent spectra for a given set of atmospheric and physical parameters, and these model spectra are compared to the data within a Bayesian inference framework to estimate the posterior probability distribution functions (PDFs) of the model parameters. In this section, we first describe the atmospheric retrieval setup followed by how the retrieval model is used for time-series analysis.

### *3.1. Atmospheric Model Setup*

We divide the atmosphere into 64 equally spaced layers with pressure levels ranging from $10^{-3}$ to $10^{2.3}$ bar (or 0.001 to 199.52 bar). The atmosphere is assumed to be cloud-free, with continuum opacity contributed by $H_2$ and He via collision-induced absorption (e.g., $H_2$–$H_2$, $H_2$–He). The remaining opacity sources are assumed to arise from $H_2O$, $CH_4$, $CH_3D$, $^{12}$CO,[16] $CO_2$, $NH_3$, $H_2S$, and $PH_3$. The volume mixing ratios (VMRs) of these gases are treated as free parameters and are assumed to be constant throughout the atmosphere (i.e., uniform with pressure). The physical properties, including mass ($M$) and radius ($R$), are also treated as free parameters and are used to determine the surface gravity.

[16] Hereafter, CO denotes $^{12}$CO unless explicitly stated otherwise.

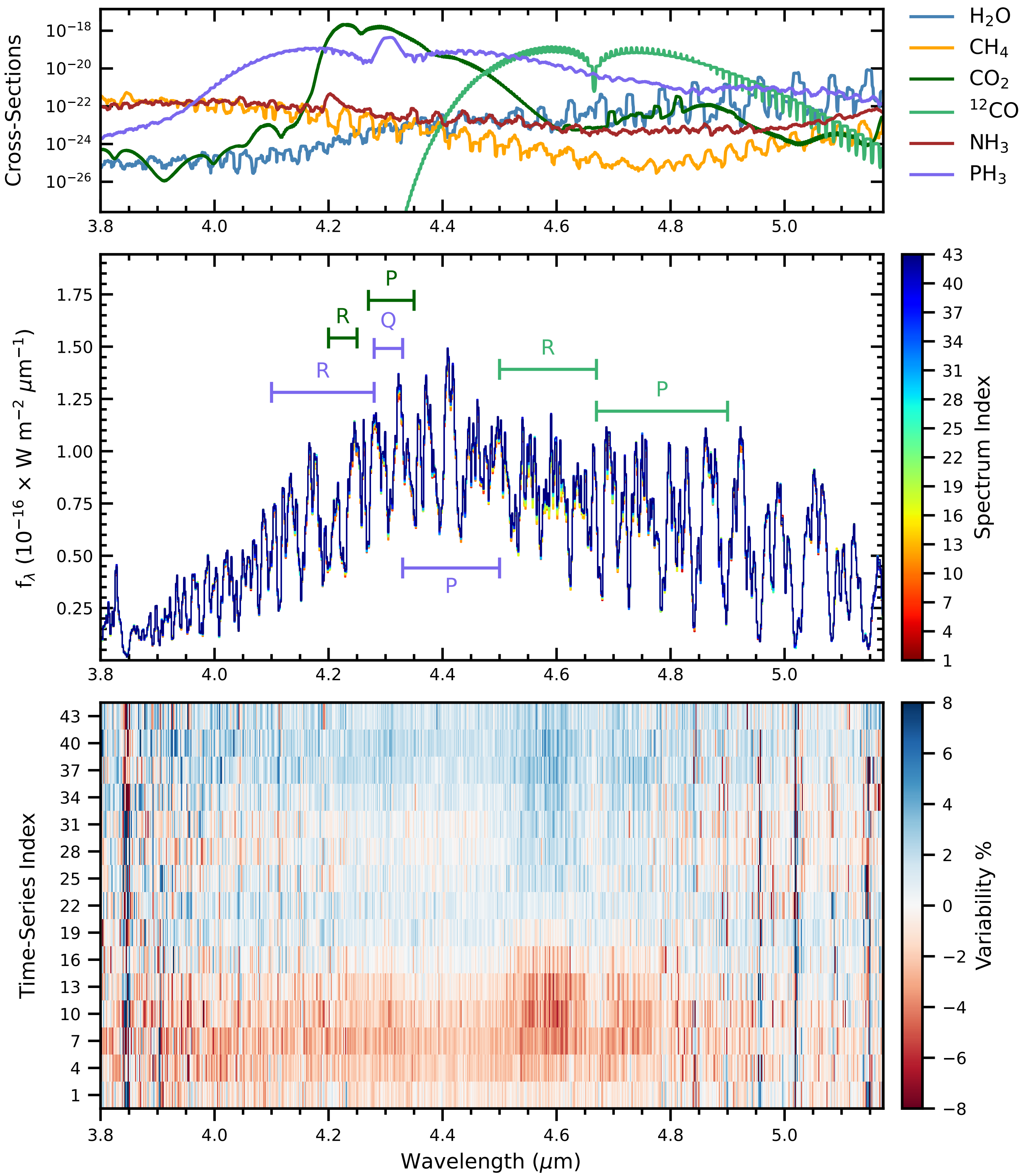

**Figure 1.** The top panel shows molecular absorption cross sections at 1 bar and 500 K for the major species in WISE 0855−07. The middle panel shows every third spectrum from the time-series dataset used for retrieval analysis, with P-, Q-, and R-branch structures indicated for key variable absorbers to highlight their correspondence with molecular opacity features. Colors for each molecule are consistent with those shown in the top panel. The bottom panel displays the spectral variability percent map relative to the overall mean spectrum. The most prominent variability aligns with $^{12}CO$ near 4.6–4.9 $\mu$m, $PH_3$ near 4.1–4.5 $\mu$m, and $CO_2$ near 4.26 $\mu$m.

The thermal structure is parameterized using a nine-knot spline penalizing interpolation. The placement of the knots is informed by an initial retrieval test on the mean spectrum of the time-series dataset, in which we first adopted equally spaced pressure knots. Based on the resulting contribution functions (shown in Figure 17; see also Section 4.1), we then placed the knots approximately equidistant within three atmospheric regions: top [0.001, 0.007, 0.058] bar, middle [0.45, 1.23, 3.41, 9.44] bar, and bottom [43.40, 199.52] bar.

### 3.2. Justification for a Cloud- and Condensation-free Atmosphere

Although condensate clouds, particularly water-ice clouds, are expected to form in the atmospheres of extremely cold brown dwarfs, we adopt a cloud-free and $H_2O$ condensation-free retrieval framework for this study. In this context, "cloud-free" means that we do not include an explicit cloud opacity parameterization, and "$H_2O$ condensation-free" means that the $H_2O$ gas-phase VMR is assumed to be uniform with pressure. This choice is motivated by both the vertical sensitivity of the present dataset and the level of model complexity justified by the data.

The potential importance of water-ice clouds in WISE 0855 −07 has been discussed extensively in the literature. Atmospheric models predict that, at the temperatures relevant for this object, $H_2O$ can condense in the upper atmosphere and form water-ice clouds that may affect the emergent spectrum (C. V. Morley et al. 2014a, 2014b). Observations at 4.5–5.2 $\mu$m have revealed $H_2O$ vapor absorption in WISE 0855 −07 and have been interpreted as broadly consistent with a Jupiter-like atmosphere in which clouds may contribute to the spectral appearance (A. J. Skemer et al. 2016; C. V. Morley et al. 2018). More recently, H. Kühnle et al. (2025) performed a joint atmospheric retrieval analysis of WISE 0855−07 using NIRSpec/PRISM, NIRSpec/G395M (K. L. Luhman et al. 2024), and MIRI/MRS spectra, and found evidence for depletion of $H_2O$ in the upper atmosphere consistent with the onset of condensation. However, despite this evidence for atmospheric $H_2O$ condensation, they did not identify direct spectral signatures attributable to water-ice cloud features across the combined near- to mid-infrared wavelength range (H. Kühnle et al. 2025).

Our decision to not include clouds in the retrieval is based on the limited vertical sensitivity of the JWST/NIRSpec G395M dataset to the atmospheric layers where water-ice clouds are expected to reside. The contribution functions for our spectra indicate that the strongest sensitivity is concentrated at pressures of approximately 0.3–10 bar (Figure 5), whereas $H_2O$ condensation and associated cloud formation are expected to occur at lower pressures ($\sim$ 0.1 bar) and colder regions of the atmosphere (C. V. Morley et al. 2014b). As a result, cloud properties such as cloud-top pressure, particle size, and optical depth are only weakly constrained by the wavelengths and pressure levels probed by these data. Including such parameters would therefore primarily introduce additional degeneracies with the thermal profile and molecular abundances rather than provide robust constraints.

More generally, in atmospheric retrievals, additional parameters are only warranted when the data require them. Introducing cloud parameters increases model flexibility, but Bayesian inference naturally penalizes unnecessary complexity through the marginal likelihood, which encodes the trade-off between improved fit quality and expanded prior volume (R. Trotta 2008). Since our cloud-free model already provides an adequate description of the spectra, we do not introduce additional cloud parameters that are unlikely to be meaningfully constrained by the present dataset.

This interpretation is also consistent with the mean spectrum retrieval analysis of M. J. Rowland et al. (2024), who found that, using the G395M spectrum alone, Bayesian information criterion (BIC) comparisons did not favor a nonuniform $H_2O$ VMR profile. Although the more flexible profile provided a better fit, the improvement was not sufficient to justify the introduction of two additional parameters describing vertical variation in the $H_2O$ VMR. This result further supports the conclusion that the present G395M dataset does not strongly warrant added vertical complexity in the atmospheric model.

This modeling choice is also consistent with recent work on WISE 0855−07 by H. Kühnle et al. (2025). They found that the evidence for upper-atmospheric $H_2O$ depletion is stronger than the evidence for spectrally detectable water-ice clouds. Even when cloudy retrievals were considered, the inferred cloud layer was located deep in the atmosphere, at temperatures too high for a physically plausible water-ice cloud deck, suggesting that the retrieved cloud opacity may instead reflect compensation for missing deep opacity, an alternative thermal structure, or limitations in the adopted cloud parameterization. We therefore treat a cloud-free atmosphere as the most appropriate baseline model for the present NIRSpec G395M time-series analysis.

We note, however, that this choice does not rule out the physical presence of clouds in WISE 0855−07. A more definitive assessment will likely require broader wavelength coverage, particularly in the mid-infrared, where the colder upper atmosphere and the effects of condensation are more directly probed. Our approach is therefore best understood as a data-justified simplification: given the spectral range and pressure sensitivity of the present observations, a cloud-free framework is sufficient for interpreting the NIRSpec G395M spectra and their time-dependent variability.

### 3.3. Opacity Treatment

For the gaseous opacities, we precompute linelist cross sections over our full temperature–pressure ($T$–$P$) domain using the following line lists: $H_2O$ (O. L. Polyansky et al. 2018), $CH_4$ (R. J. Hargreaves et al. 2020), $CH_3D$ (R. J. Hargreaves et al. 2020), $^{12}CO$ (L. Rothman et al. 2010; G. Li et al. 2015), $CO_2$ (X. Huang et al. 2014), $NH_3$ (S. N. Yurchenko & J. Tennyson 2014), $H_2S$ (A. A. A. Azzam et al. 2015), and $PH_3$ (C. Sousa-Silva et al. 2015). These cross sections are tabulated at fixed intervals of 0.5 dex in pressure with temperature ranging from 75 up to 4000 K, with temperature intervals ranging from 25 to 250 K depending on the temperature regime. For intermediate $T$–$P$ points, we obtain the opacities via linear interpolation in temperature space and linear in $\log_{10}$(P) space.

In addition to molecular and atomic line opacities adopted from L. S. Rothman et al. (2010), we include continuum opacity sources from collisionally induced absorption (CIA) of $H_2$–$H_2$ and $H_2$–He from C. Richard et al. (2012) and D. Saumon et al. (2012).

The final noncontinuum cross sections span our full observed wavelength range of ~2.8–5.2 $\mu$m and are computed at a piecewise linearly varying resolving power ($R = \lambda/\Delta\lambda$). Specifically, the resolving power increases from $R \approx$ 40,000 at 2.8 $\mu$m to $R \approx$ 60,000 at 3.7 $\mu$m, and continues increasing to $R \approx$ 100,000 at 5.2 $\mu$m, ensuring that the cross sections are sampled at or above the local instrumental resolution across the full wavelength range.

### 3.4. Generative Model

Given a set of atmospheric parameters, the forward model computes the emergent spectrum at the top of the atmosphere using the two-stream source function technique described in O. B. Toon et al. (1989). The model-predicted flux at each wavelength, $\mathcal{M}_\lambda(\lambda_i)$, is then given by

$$\mathcal{M}_\lambda(\lambda_i) = \left(\frac{R}{d}\right)^2 [\boldsymbol{I}(\lambda_i) * \mathcal{F}_\lambda(\boldsymbol{\theta}_{\rm atm}, \lambda_j)], \quad (3)$$

where

1. $\mathcal{F}_\lambda(\boldsymbol{\theta}_{\rm atm}, \lambda_j)$ is the model emergent flux densities at the top of the atmosphere given a set of atmospheric parameters ($\boldsymbol{\theta}_{\rm atm}$), calculated using a two-stream source function technique described in O. B. Toon et al. (1989). $\lambda_j$ is equal to $\lambda_k + \Delta\lambda$, where $\lambda_k$ is the wavelength at which the model emergent flux density is calculated and $\Delta\lambda$ is a parameter that accounts for uncertainty in wavelength.
2. $\boldsymbol{I}(\lambda_i)$ is the instrument profile, modeled as Gaussian, which accounts for the variable resolving power of the observed spectrum at each wavelength ($\lambda_i$).
3. $R / d$ is a scaling factor to scale the model spectrum as it is observed from Earth.

Each datum in the observed spectrum is modeled probabilistically as

$$F_\lambda(\lambda_i) = \mathcal{M}_\lambda(\lambda_i) + \epsilon(\lambda_i), \quad (4)$$

where $F_\lambda(\lambda_i)$ is a random variable denoting the flux density at wavelength $\lambda_i$ and $\epsilon(\lambda_i)$ is a Gaussian random parameter with zero mean and variance given by $\sigma^2(\lambda_i)$.

### 3.5. Retrieval Framework and Likelihood Function

The forward model described above is used by the `Brewster` retrieval framework (B. Burningham et al. 2017) to infer the atmospheric properties of WISE 0855−07 from the observed spectra. The parameter space is explored using the nested sampling algorithm PyMultiNest (J. Buchner et al. 2014), a Python implementation of MultiNest (F. Feroz et al. 2009). For all retrievals reported in this work, we used 500 live points, with constant efficiency mode and importance nested sampling disabled. This setup was adopted for consistency across all retrieval scenarios and to provide a computationally tractable sampling configuration for the large number of retrievals performed in this analysis. For this work, `Brewster` has been extended to perform time-varying spectral retrievals, fitting each epoch in the time-series dataset independently while using a common forward-model framework.

The model parameter vector $\boldsymbol{\Theta}$ includes the gas VMRs, the thermal profile parameters, the physical parameters such as mass and radius, and nuisance parameters (see Table 1). The posterior probability of the model parameters given the observed spectrum $\boldsymbol{f}_\lambda$ follows from Bayes' theorem

$$p(\boldsymbol{\Theta}|\boldsymbol{f}_\lambda) = \frac{p(\boldsymbol{\Theta})\mathcal{L}(\boldsymbol{f}_\lambda|\boldsymbol{\Theta})}{p(\boldsymbol{f}_\lambda)}. \quad (5)$$

where $p(\boldsymbol{\Theta}|\boldsymbol{f}_\lambda)$ is the posterior PDF, $p(\boldsymbol{\Theta})$ is the prior PDF, $\mathcal{L}(\boldsymbol{f}_\lambda|\boldsymbol{\Theta})$ is the likelihood function, and $p(\boldsymbol{f}_\lambda)$ is the Bayesian evidence, which can be used for model comparison.

**Table 1**
Parameter Priors

| Parameter | Description | Prior[a] |
|---|---|---|
| $\log_{10}(f_i)$[b,c] | Gas Volume Mixing Ratio | $\mathcal{U}(-12, 0)$ |
| $M$ | Mass ($\mathcal{M}^{\rm N}_{\rm Jup}$) | $\mathcal{U}(1, 80)$ |
| $R$ | Radius ($\mathcal{R}^{\rm N}_{e\rm J}$) | $\mathcal{U}(0.5, 2)$ |
| $\Delta\lambda$ | Wavelength shift ($\mu$m) | $\mathcal{U}(-0.01, 0.01)$ |
| $10^b$ | Error inflation | $\mathcal{U}(0.01 \times \min(\sigma_i^2), 100 \times \max(\sigma_i^2))$ |
| $d$ | Distance (pc)[d] | $\mathcal{N}(\mu, \sigma^2)$ |
| $T_{\rm Knot\text{-}i}$ | Thermal profile knots (K) | $\mathcal{U}(0, 5000)$ |
| $\gamma$ | Thermal profile smoothing hyperparameter | $\mathcal{U}(0, 10000)$ |

**Notes.**
[a] $\mathcal{U}(\alpha, \beta)$ denotes a uniform distribution between $\alpha$ and $\beta$ while $\mathcal{N}(\mu, \sigma^2)$ denotes a normal distribution with a mean of $\mu$ and a variance of $\sigma^2$.
[b] Our retrieved model atmospheric gases include $H_2O$, $CH_4$, $CH_3D$, $^{12}CO$, $CO_2$, $NH_3$, $H_2S$, and $PH_3$.
[c] All VMRs (the number density of the species divided by the total number density of the gas) are reported as the log of the ratio, and the remainder of the gas is assumed to be $H_2$–He ($1-\sum_i f_i$). Assuming a solar abundance of 91.2% of the number of atoms of H and 8.7% of the number of atoms of He (M. Asplund et al. 2009), 84% of the VMR is from $H_2$ and 16% is from He for the remainder of the gas.
[d] The prior for distance was adapted from J. D. Kirkpatrick et al. (2021).

Our full log-likelihood function consists of two components

$$\ln \mathcal{L}_{\rm total} = \ln \mathcal{L} + \ln p(\boldsymbol{T}, \gamma), \quad (6)$$

where the first term is the data–model likelihood and the second term is the penalizing prior on the thermal profile, described below.

Assuming the noise in the observed spectrum is Gaussian and the spectral points are independent, the data–model log-likelihood is:

$$\ln\mathcal{L}(\boldsymbol{f}_\lambda|\boldsymbol{\Theta}) = -\frac{1}{2}\sum_{i=1}^{n}\left\{\frac{[f_{\lambda,i} - \mathcal{M}_\lambda(\lambda_i)]^2}{\sigma^2_{\rm model}(\lambda_i)} + \ln[2\pi\sigma^2_{\rm model}(\lambda_i)]\right\}, \quad (7)$$

where $f_{\lambda,i}$ is the observed flux at wavelength $\lambda_i$, $\mathcal{M}_\lambda(\lambda_i)$ is the predicted model flux described in Section 3, and $\sigma^2(\lambda_i)$ is the total variance at $\lambda_i$, modeled as

$$\sigma^2_{\rm model}(\lambda_i) = s^2(\lambda_i) + 10^b, \quad (8)$$

where $s(\lambda_i)$ is the standard error of the measured flux at $\lambda_i$, and $b$ is a tolerance parameter to account for unmodeled systematic uncertainties (e.g., D. W. Hogg et al. 2010; D. Foreman-Mackey et al. 2013; B. Burningham et al. 2017).

We adopt the penalizing thermal profile parameterization from M. R. Line et al. (2015), which is flexible enough to capture thermal structure while discouraging nonphysical oscillations. The prior on the thermal profile $p(\boldsymbol{T}, \gamma)$ penalizes roughness via the discrete second derivative of the temperature structure across pressure knots. Because the likelihood function and $p(\boldsymbol{T}, \gamma)$ are both evaluated in log space, and because there is no straightforward way to penalize the prior probability directly within nested sampling, we incorporate this penalty into the total log-likelihood (Equation (6)) as

$$\ln p(\boldsymbol{T}, \gamma) = -\frac{1}{2\gamma}\sum_{i=0}^{N}(T_{i+1} - 2T_i + T_{i-1})^2 - \frac{1}{2}\ln(2\pi\gamma), \quad (9)$$

where the summation involves the discrete second derivative of the temperature structure at each pressure knot $i$, which quantifies the local curvature of the thermal profile. The contribution of these second-derivative terms is weighted by the parameter $\gamma$, which controls the strength of the penalty on roughness. Large values of this term indicate sharp changes in slope (i.e., roughness), while values close to zero correspond to smoother, more linear behavior across adjacent pressure knots. Thus, $\gamma$ effectively smooths the thermal profile by discouraging abrupt variations, without enforcing a fixed functional form for the temperature structure. For flexibility, $\gamma$ is treated as a free parameter drawn from an inverse gamma distribution ($\tilde{\Gamma}$), with the hyperpriors shown in Table 1. This parameterization has been previously used in M. R. Line et al. (2015, 2017) and J. A. Zalesky et al. (2019).

### 3.6. Fiducial Retrieval Setup

We begin with a fiducial retrieval performed on the first spectrum in the time-series dataset, which serves as the baseline for all subsequent variability models.

Using the model setup described above, we carried out a full atmospheric retrieval on the first spectrum from the time-series dataset and tested three different equivalent resolving powers for the line-list cross sections: constant $R = 40{,}000$, variable $R = 40{,}000$–$100{,}000$, and variable $R = 60{,}000$–$100{,}000$. For all subsequent analyses, we adopted the variable $R = 40{,}000$–$100{,}000$ equivalent resolving power. The retrieved parameter values from this setup were statistically consistent, within $1\sigma$, with those obtained using the variable $R = 60{,}000$–$100{,}000$ case, but at substantially lower computational cost. In contrast, the constant $R = 40{,}000$ case produced systematically higher gas VMRs, with deviations exceeding $1\sigma$.

### 3.7. Quantifying the Goodness of Fit of Variability Maps

To assess how well a given retrieval model reproduces the observed time-dependent variability, we compare the observed and model variability maps in wavelength–time-index space. Figures 2 and 3 illustrate the basic framework. Figure 2 shows that the observed variability map contains both coherent astrophysical signal and stochastic noise, while Figure 3 shows the idealized retrieval outcome in which the model captures the structured variability and the residual map is dominated by noise. We quantify the goodness of fit using the noise-weighted rms residual variability, which measures the typical unexplained variability amplitude remaining after subtracting the model from the data.

#### 3.7.1. Normalized Flux Fields and Residual Variability

For each spectrum index $i$ and wavelength bin $\lambda$, we define the observed and model normalized flux fields as

$$\delta_{i,\lambda}^{\rm obs} = \frac{F_{i,\lambda}^{\rm obs}}{\bar{F}_{\lambda}^{\rm obs}}, \qquad \delta_{i,\lambda}^{\rm mod} = \frac{F_{i,\lambda}^{\rm mod}}{\bar{F}_{\lambda}^{\rm mod}}, \quad (10)$$

where the mean spectrum at each wavelength is

$$\bar{F}_{\lambda}^{\rm obs} = \frac{1}{N_t}\sum_{i=1}^{N_t} F_{i,\lambda}^{\rm obs}, \qquad \bar{F}_{\lambda}^{\rm mod} = \frac{1}{N_t}\sum_{i=1}^{N_t} F_{i,\lambda}^{\rm mod}. \quad (11)$$

The variability maps shown in this work are constructed from these normalized quantities and expressed as the percent deviation from the observed mean spectrum

$$V_{i,\lambda}^{\rm obs} = 100 \times (1 - \delta_{i,\lambda}^{\rm obs}), \quad (12)$$

and similarly from the model mean spectrum

$$V_{i,\lambda}^{\rm mod} = 100 \times (1 - \delta_{i,\lambda}^{\rm mod}). \quad (13)$$

For goodness-of-fit calculations, we work directly with the residual variability field in fractional units

$$r_{i,\lambda} = \delta_{i,\lambda}^{\rm obs} - \delta_{i,\lambda}^{\rm mod}. \quad (14)$$

The corresponding residual variability map in percent is then

$$V_{i,\lambda}^{\rm res} = 100 \times r_{i,\lambda}. \quad (15)$$

This definition is consistent with the schematic in Figure 3, where the residual map represents the variability remaining after subtracting the retrieved model from the observed variability map, which looks like the noise variability from Figure 2.

#### 3.7.2. Propagation of Uncertainties

Let $\sigma_{i,\lambda}$ denote the measured $1\sigma$ observational uncertainty on the flux density $F_{i,\lambda}^{\rm obs}$. For this diagnostic calculation, we use the measured observational uncertainties, rather than the retrieval tolerance term introduced in Equation (8), to define the weights. This choice keeps the residual variability metric tied to the data precision and avoids introducing a model-dependent noise floor when comparing different variability retrieval scenarios. Since the observed normalized flux is obtained by dividing by the mean observed spectrum, the propagated uncertainty in $\delta_{i,\lambda}^{\rm obs}$ is approximately

$$\sigma_{\delta,i,\lambda} \approx \frac{\sigma_{i,\lambda}}{\bar{F}_{\lambda}^{\rm obs}}, \quad (16)$$

where we neglect the comparatively small uncertainty in the time-averaged mean spectrum. Because the model is treated as deterministic for a given retrieved parameter set, the uncertainty in the residual field is approximated as

$$\sigma_{r,i,\lambda} \approx \sigma_{\delta,i,\lambda}. \quad (17)$$

#### 3.7.3. Noise-weighted rms Residual Variability

We define the noise-weighted rms residual variability as

$$\mathrm{wRMS}_{\rm res} = \sqrt{\frac{\sum_{i=1}^{N_t}\sum_{\lambda=1}^{N_\lambda} w_{i,\lambda}\, r_{i,\lambda}^2}{\sum_{i=1}^{N_t}\sum_{\lambda=1}^{N_\lambda} w_{i,\lambda}}}, \quad (18)$$

with inverse-variance weights

$$w_{i,\lambda} = \frac{1}{\sigma_{r,i,\lambda}^2}. \quad (19)$$

To report this quantity on the same scale as the plotted variability maps, we convert it to percent:

$$\mathrm{wRMS}_{\rm res}^{(\%)} = 100 \times \mathrm{wRMS}_{\rm res}. \quad (20)$$

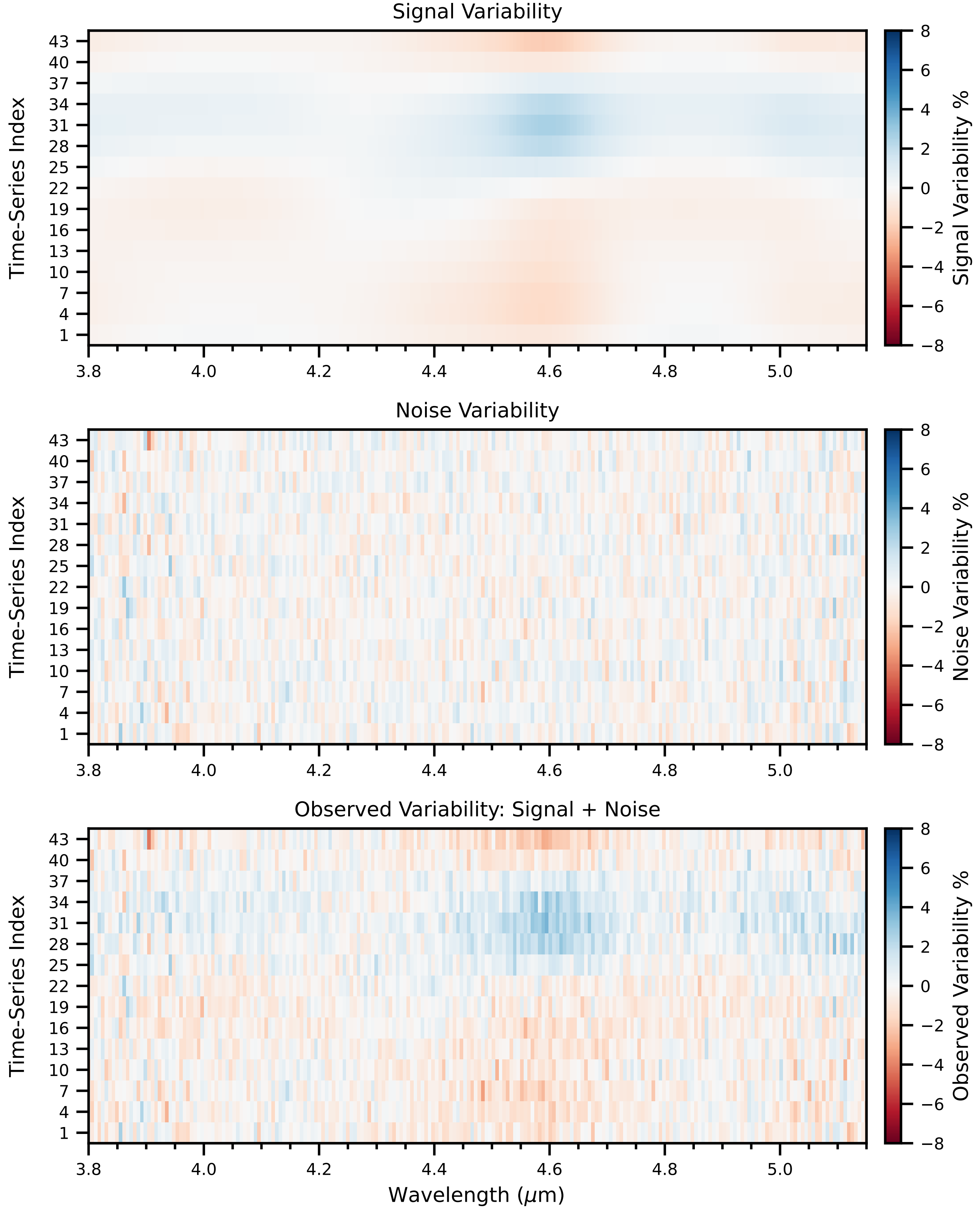

**Figure 2.** The top panel shows an idealized astrophysical signal variability map, the middle panel shows the corresponding noise map, and the bottom panel shows the observed variability map formed by their sum. The signal and the noise are normalized by their respective mean values.

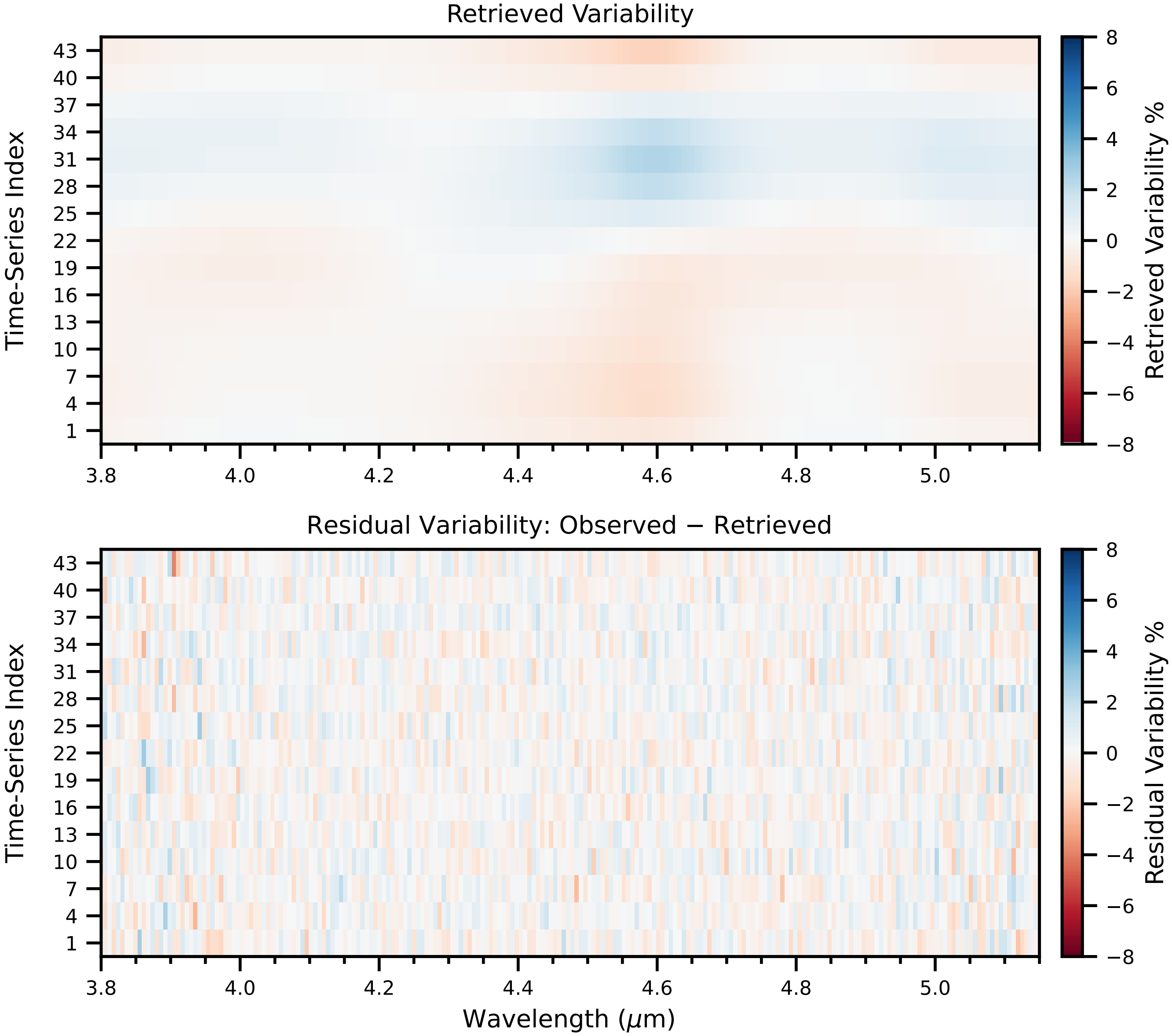

**Figure 3.** The top panel shows the variability recovered by the retrieval model, while the bottom panel shows the residual variability map obtained by subtracting the retrieved model from the observed variability map. In the ideal case, the retrieval captures the coherent signal present in the data, leaving residuals that are dominated by noise. We quantify the quality of the fit using the weighted rms residual variability, for which smaller values indicate that less structured variability remains unexplained by the retrieval model.

This statistic represents the typical unexplained time-dependent variability amplitude after accounting for the wavelength-dependent measurement uncertainties.

*3.7.4. Noise Floor and Interpretation*

To determine whether the residual variability is consistent with the measurement precision, we compute the corresponding noise floor in the same normalization

$$\text{Noise}\quad\text{Floor} = 100 \times \sqrt{\frac{\sum_{i=1}^{N_t}\sum_{\lambda=1}^{N_\lambda} w_{i,\lambda}\, \sigma_{r,i,\lambda}^2}{\sum_{i=1}^{N_t}\sum_{\lambda=1}^{N_\lambda} w_{i,\lambda}}}. \tag{21}$$

Because $w_{i,\lambda} = 1/\sigma_{r,i,\lambda}^2$, this expression reduces to the weighted rms of the expected noise in the residual map. The comparison between $\text{wRMS}_{\text{res}}^{(\%)}$ and the noise floor has a straightforward interpretation:

1. If $\text{wRMS}_{\text{res}}^{(\%)} >$ noise floor, then structured variability remains in the residual map, implying that the model does not fully capture the observed variability.
2. If $\text{wRMS}_{\text{res}}^{(\%)} \approx$ noise floor, then the residual variability is consistent with the expected measurement noise, implying that no substantial coherent variability remains unexplained.

*3.7.5. Use in Comparing Variability Models*

This metric is well suited for comparing competing variability prescriptions because it compresses the full two-dimensional wavelength–time residual structure into a single physically interpretable quantity while naturally down-weighting noisier spectral regions. In practice, smaller values of $\text{wRMS}_{\text{res}}^{(\%)}$ indicate that less coherent variability remains unexplained, and values approaching the noise floor indicate that the remaining residual structure is predominantly noise-like. This provides a straightforward and statistically meaningful basis for comparing retrieval models for time-dependent spectral variability.

## 4. Results

In this section, we present the results of our time-variable retrieval analysis. We first describe the fiducial retrieval performed on the first spectrum in the time-series dataset, which provides the baseline model for all subsequent variability experiments. This fiducial retrieval yields a 23-dimensional posterior PDF, from which we derive marginalized 1D posterior distributions for each parameter (see Figure A1).

Using the fiducial case as a reference, we evaluate variability models on every third spectrum in the time series, corresponding to 14 spectra in total. Each time-series spectrum is fitted independently. In a given variability scenario, only the parameters associated with that scenario are allowed to vary from epoch to epoch, while all remaining atmospheric and physical parameters are fixed to the posterior median values from the fiducial retrieval. The priors on the free parameters in the variability retrievals are the same as those used in the fiducial retrieval and are listed in Table 1.

We present the variability scenarios in order of decreasing residual structure, from the least successful to the most successful model. We consider a total of seven variability model scenarios, grouped into three categories:

1. VMR-only models: CO, CO + $PH_3$, and CO + $CO_2$ + $PH_3$.
2. Temperature-only model.
3. Temperature + VMR models: Temperature + CO, Temperature + CO + $PH_3$, and Temperature + CO + $CO_2$ + $PH_3$.

This model hierarchy is motivated by the wavelength-dependent variability shown in Figure 1. As discussed in Section 2, the largest peak-to-peak variations occur across the CO fundamental band at 4.5–4.9 $\mu$m and near the $CO_2\nu_3$ band at 4.26 $\mu$m, with additional modulation over 4.1–4.5 $\mu$m where $PH_3\nu_2$ absorption overlaps the CO P-branch. These features motivate a set of VMR-only variability models in which the abundances of CO, $PH_3$, and $CO_2$ are allowed to vary across epochs to test whether composition changes alone can reproduce the observed variability pattern.

We next consider a temperature-only model to test the complementary hypothesis that variability is driven by changes in the thermal structure, which can produce wavelength-dependent changes in emergent flux across molecular bands as shown in E. Nasedkin et al. (2025). Finally, because thermal perturbations and VMR variations may coexist in a dynamically active atmosphere, we also explore temperature + VMR variability models in which both are allowed to vary simultaneously. This framework allows us to incrementally assess the potential driver(s) of the observed variability.

### 4.1. Fiducial Retrieval

The top panel of Figure 4 compares the observed spectrum to the retrieved model spectrum calculated using the posterior median parameter values (hereafter the median model spectrum). The shaded regions around the median model spectrum denote the $2\sigma$ central credible intervals.[17] The bottom panel shows the residual between the observed spectrum and the median model spectrum. Overall, the model provides a good fit to the data, with small mismatches across 3.8–5.2 $\mu$m, which likely arise from an underestimation of model uncertainties.

Figure 5 shows the retrieved fiducial thermal profile. The black curve indicates the posterior median profile, while the shaded blue regions denote the $1\sigma$ and $2\sigma$ central credible intervals. The gray dashed curve shows the normalized mean contribution function (J. W. Chamberlain & D. M. Hunten 1987), highlighting the pressure levels most strongly probed by the observations. The overlap between the dominant region of the mean contribution function and the narrowing of the credible intervals indicates the atmospheric layers most strongly constrained by the first observed spectrum, and by extension, the dataset as a whole.

For comparison, the red curve shows the thermal profile from the best-fit coolTLUSTY forward model from Miles et al. (2026, submitted), which is a cloudy and disequilibrium chemistry treatment, with $T_{\rm eff} = 250$ K, CDAMPU = 6 (this parameter sets the ratio of gaseous materials' pressure scale height to the cloud particles pressure scale height), $\log_{10}(K_{zz}) = 6$ cm$^2$ s$^{-1}$, $\log_{10}(g) = 4$ cm s$^{-2}$, and a radius of 1 $R_{\rm Jup}$. For more details about the model, please refer to Miles et al. (2026, submitted). This forward model provides an independent, physically motivated atmospheric structure against which the retrieved profile can be compared. Over the pressures most strongly probed by the observations, represented by the mean contribution function in gray, the forward-model profile is broadly consistent with the retrieved median profile. However, at higher pressures, the coolTLUSTY profile becomes significantly warmer than the retrieved median profile. These deeper layers are less directly constrained by the data, as indicated by the weaker contribution function and broader posterior uncertainty, so differences between the retrieved and forward-model profiles in this region should be interpreted with caution.

### 4.2. VMR-only Variability

To guide the interpretation of the model hierarchy presented below, we order the variability scenarios according to their weighted rms residual variability, from largest to smallest. In this ordering, the models discussed first have the largest unexplained residual variability and therefore provide the least successful descriptions of the data, whereas the models discussed later yield progressively smaller residuals and better reproduce the observed wavelength-dependent variability pattern. The weighted rms residual variability values quoted throughout this section are summarized in Figure 13 and provide a quantitative complement to the variability maps discussed for each model variability scenario.

In this subsection, we examine models in which variability is introduced only through changes in molecular VMRs, while the thermal structure and all other atmospheric parameters are fixed to their fiducial median values. These scenarios test whether the observed spectral modulation can be explained by epoch-to-epoch changes in the VMR of the species most directly associated with the strongest variability features in the 4–5 $\mu$m region.

[17] A Bayesian central credible interval gives the range of values in a parameter's posterior distribution that contain $\alpha$% of the probability. In contrast, a frequentist $\alpha$% confidence interval means that $\alpha$% of a large number of confidence intervals computed in the same way would contain the true value of the parameter.

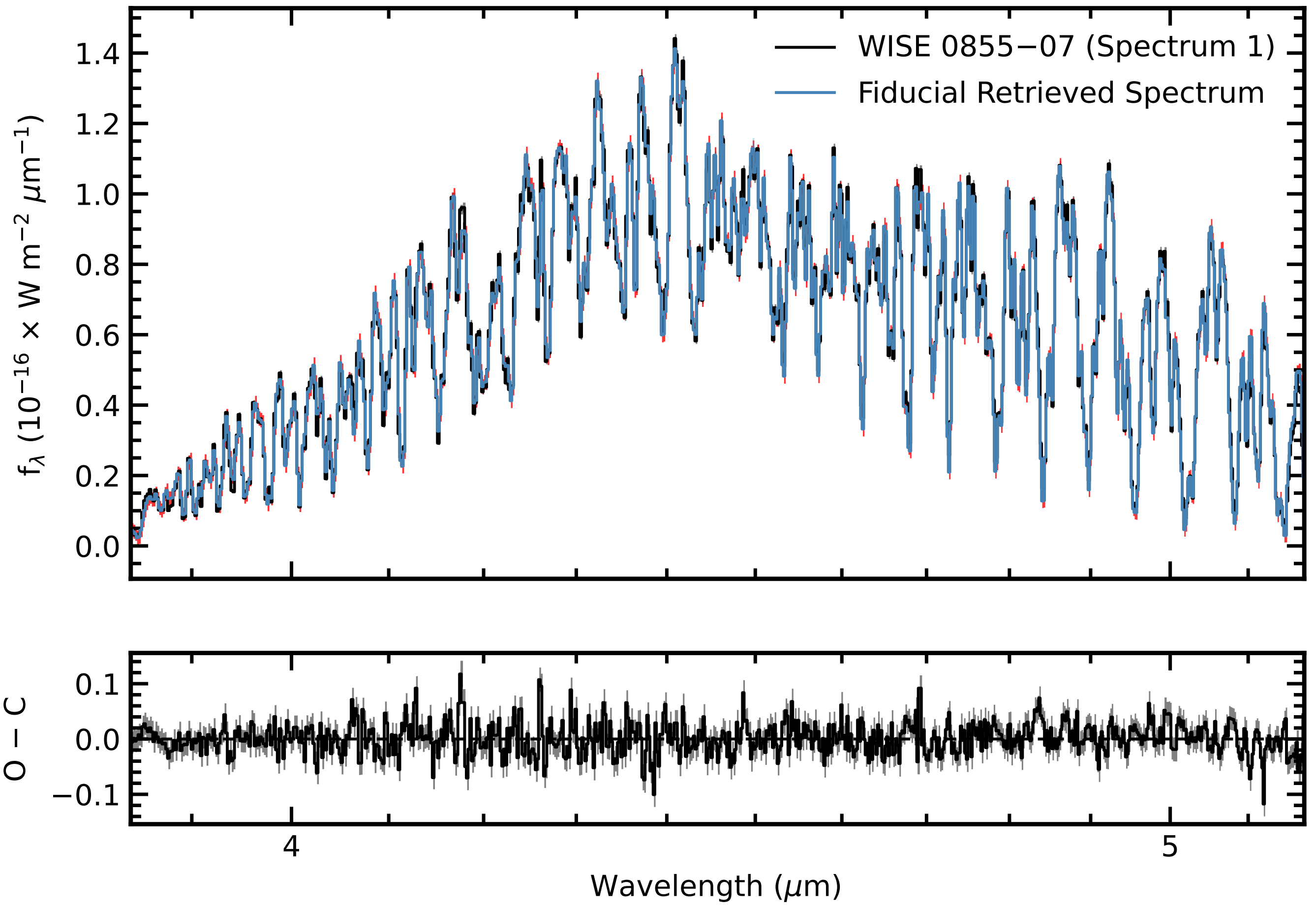

**Figure 4.** The top panel shows the comparison between the first observed spectrum in the WISE 0855−07 time-series dataset in black, and the fiducial retrieved median spectrum, shown in blue. The gray shaded region shows the 1$\sigma$ observational uncertainty, while the blue line represents the retrieved median spectrum. The red shaded region shows the total 1$\sigma$ uncertainty used in the likelihood, including both the observational uncertainty and the retrieved error-inflation term, $\sigma_{\rm model} = \sqrt{s^2 + 10^b}$ (i.e., the model uncertainty). The bottom panel shows the residuals between the observed spectrum and the fiducial retrieved median spectrum, $O - C$, with the gray shaded region indicating the corresponding propagated total 1$\sigma$ uncertainty on the residuals.

### 4.2.1. CO VMR Variability

This model isolates the extent to which CO abundance variations alone can account for the observed time-dependent spectral modulation. Figure 6 shows the variability reproduced by this model. In the top panel, the retrieved spectra are normalized by the overall mean model spectrum. The largest modeled variations occur across the CO $v = 1$–0 fundamental band at 4.5–4.9 $\mu$m, with structure concentrated in both the R-branch (4.50–4.67 $\mu$m) and P-branch (4.67–4.90 $\mu$m). This demonstrates that epoch-to-epoch changes in the CO abundance can reproduce an appreciable fraction of the observed variability within the strongest CO-dominated portion of the spectrum.

The bottom panel shows the residual variability percent between the observed and retrieved spectra, after both have been normalized by their respective mean spectra. Although the model captures part of the variability across the 4.5–4.9 $\mu$m interval, substantial residual structure remains outside this region, particularly over ∼3.8–4.5 $\mu$m and ∼4.90–5.2 $\mu$m. Consistent with this qualitative mismatch, this model yields the largest weighted rms residual variability of all scenarios considered, 1.95%, which remains well above the 1.68% noise floor.

These results indicate that CO abundance variations likely contribute to the observed variability, but are not sufficient on their own to explain the observed wavelength-dependent temporal variability pattern of WISE 0855−07.

### 4.2.2. (CO + $PH_3$) VMR Variability

We next considered a model in which the VMRs of both CO and $PH_3$ are allowed to vary. This scenario tests whether allowing two disequilibrium chemical species to vary improves the match to the observed variability pattern relative to the CO-only variability case.

Figure 7 presents the results of this model. As in the CO-only variability case, the top panel shows the retrieved spectra normalized by the overall mean model spectrum, while the bottom panel shows the residual variability percent relative to the observations. The model continues to reproduce variability across the CO fundamental band, especially within the R-branch (4.50–4.67 $\mu$m) and P-branch (4.67–4.90 $\mu$m). In addition, allowing $PH_3$ to vary introduces variability in the 4.2–4.5 $\mu$m region associated with the $PH_3$ band structure, including the R-branch (4.20–4.28 $\mu$m), Q-branch (4.28–4.33 $\mu$m), and P-branch (4.33–4.50 $\mu$m).

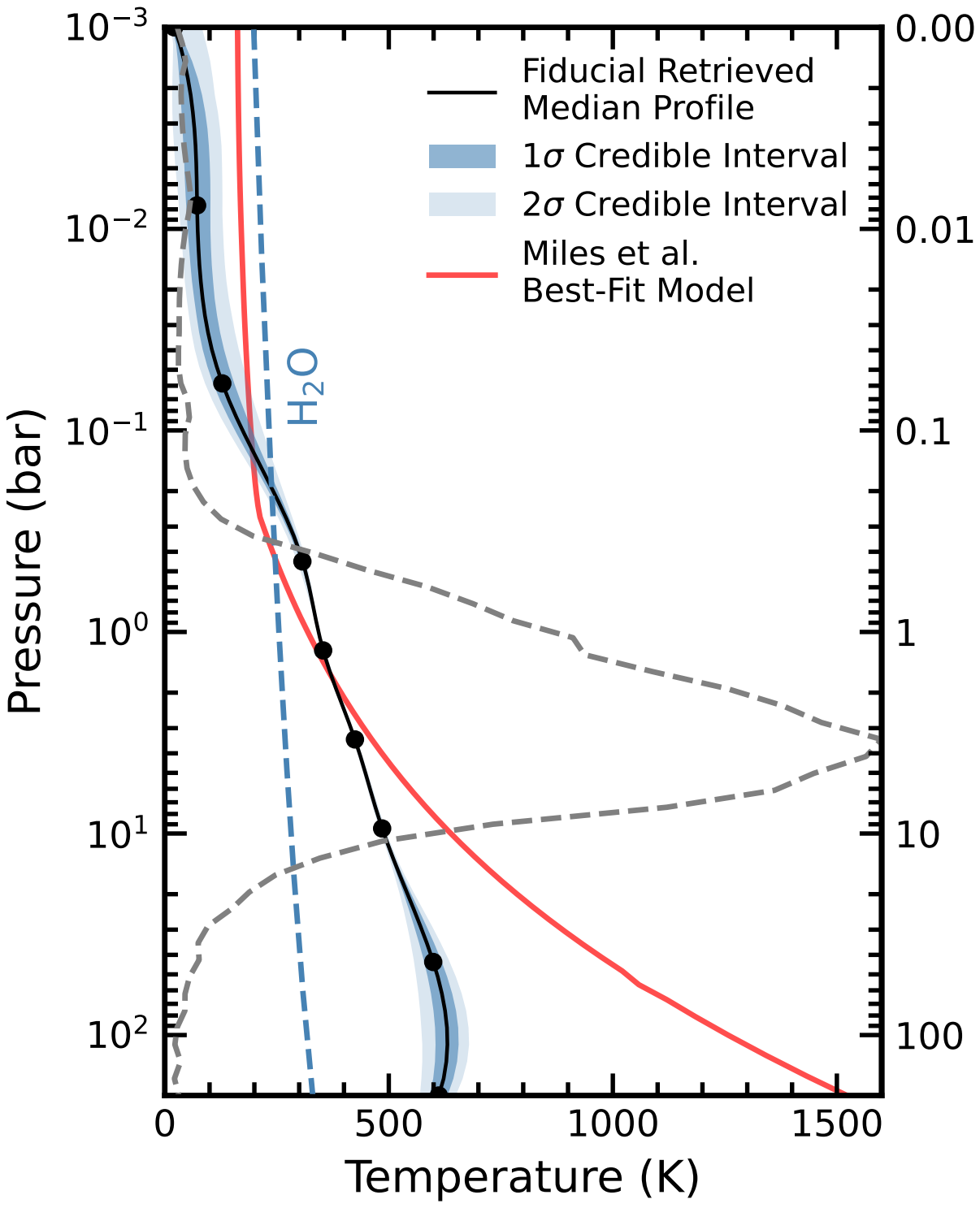

**Figure 5.** Retrieved thermal profile from the fiducial retrieval. The black line shows the retrieved median profile, while the shaded blue regions represent the $1\sigma$ and $2\sigma$ central credible intervals. The gray dashed curve indicates the mean contribution function, which highlights the pressure levels that most strongly contribute to the observed flux. The black dots represent the temperature–pressure knots. The blue dashed line represents the water condensation curve. The red curve shows the thermal profile of the best-fit coolTLUSTY forward model from B. E. Miles et al. (2026), with $T_{\rm eff} = 250$ K, CDAMPU = 6, and $\log(K_{zz}) = 6$ $\rm cm^2 s^{-1}$.

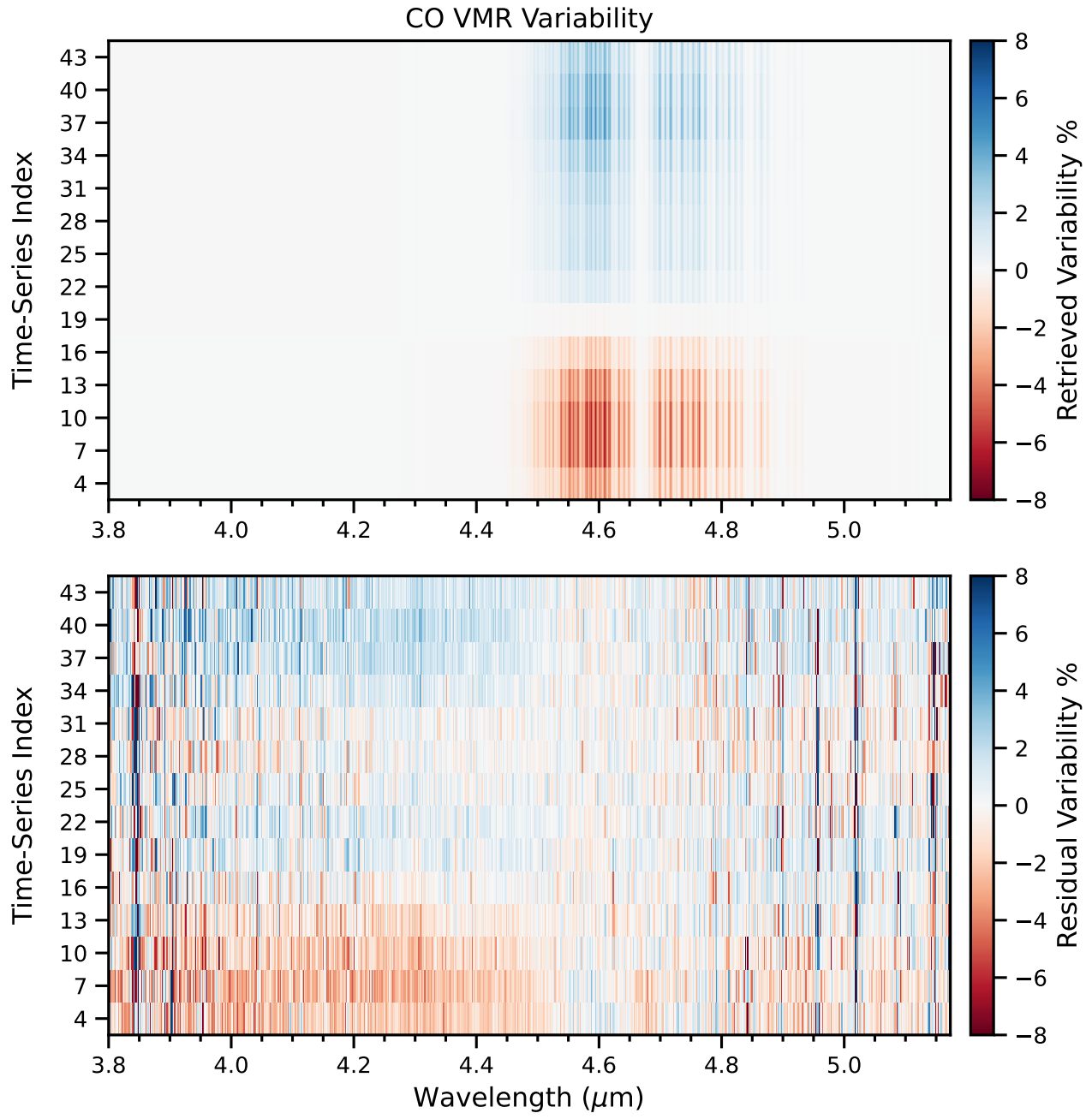

**Figure 6.** Spectral variability captured by the CO VMR variability model. The top panel shows the retrieved spectra normalized to the overall mean model spectrum, illustrating the variability reproduced when only the CO VMR is allowed to vary. The bottom panel shows the residual variability percent between the observed and retrieved spectra, with both normalized by their respective mean spectra. This model scenario yields a weighted rms residual variability of 1.95%.

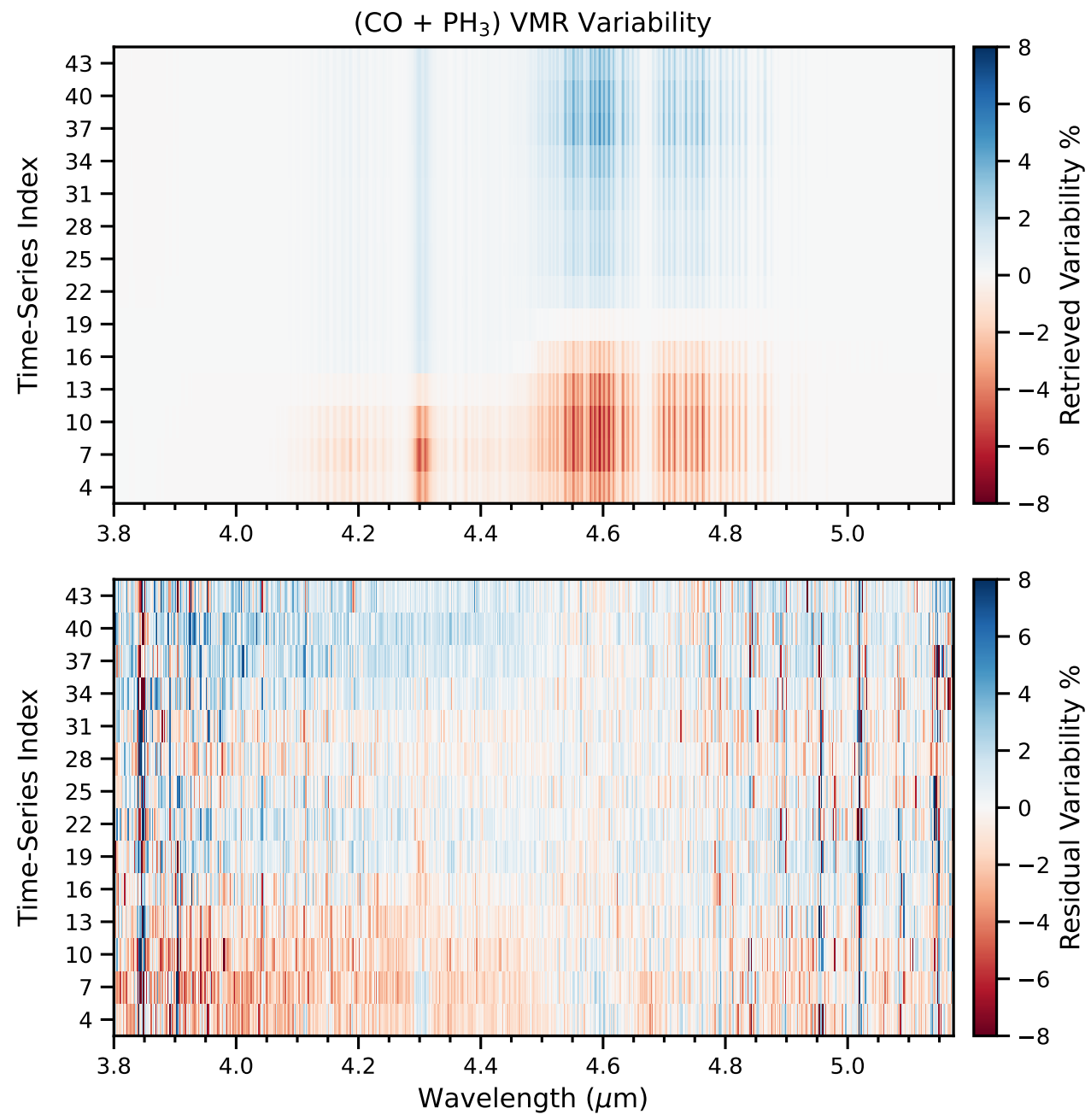

**Figure 7.** Spectral variability captured by the (CO + $PH_3$) VMR variability model. The top panel shows the retrieved spectra normalized to the overall mean model spectrum, illustrating the variability reproduced when both CO and $PH_3$ VMRs are allowed to vary. The bottom panel shows the residual variability percent between the observed and modeled spectra, with both normalized by their respective mean spectra. This model scenario yields a weighted rms residual variability of 1.87%.

Relative to the CO-only model, the inclusion of $PH_3$ reduces residual structure in parts of the 4.2–4.5 $\mu$m interval. This qualitative improvement is reflected in a lower weighted rms residual variability of 1.87%. However, this value still lies above the noise floor, and significant residuals remain across the broader 3.8–5.2 $\mu$m range, indicating that variability in CO and $PH_3$ alone does not fully reproduce the observed wavelength-dependent temporal variability.

Overall, this model supports the conclusion that abundance variations in multiple disequilibrium species contribute to the observed modulation, but that VMR variability alone is insufficient to explain the full spectral variability of WISE 0855 −07.

#### *4.2.3. (CO + $PH_3$ + $CO_2$) VMR Variability*

We then expanded the VMR-only analysis by allowing the VMR of CO, $PH_3$, and $CO_2$ to vary simultaneously, while again fixing all remaining parameters to their fiducial median values. This model tests whether the combined variability of the three species most directly associated with the observed 4–5 $\mu$m modulation can account for the measured spectral variability.

Figure 8 shows the corresponding variability map. The top panel displays the retrieved model spectra normalized by the mean model spectrum, and the bottom panel shows the residual variability percent relative to the observations. In addition to reproducing variability across the CO fundamental band and the $PH_3$ absorption region, this model yields improved agreement near 4.25 $\mu$m, where the $CO_2 \nu_3$ band contributes strongly. The reduction in residual structure in the 4.2–4.3 $\mu$m interval indicates that allowing $CO_2$ to vary helps

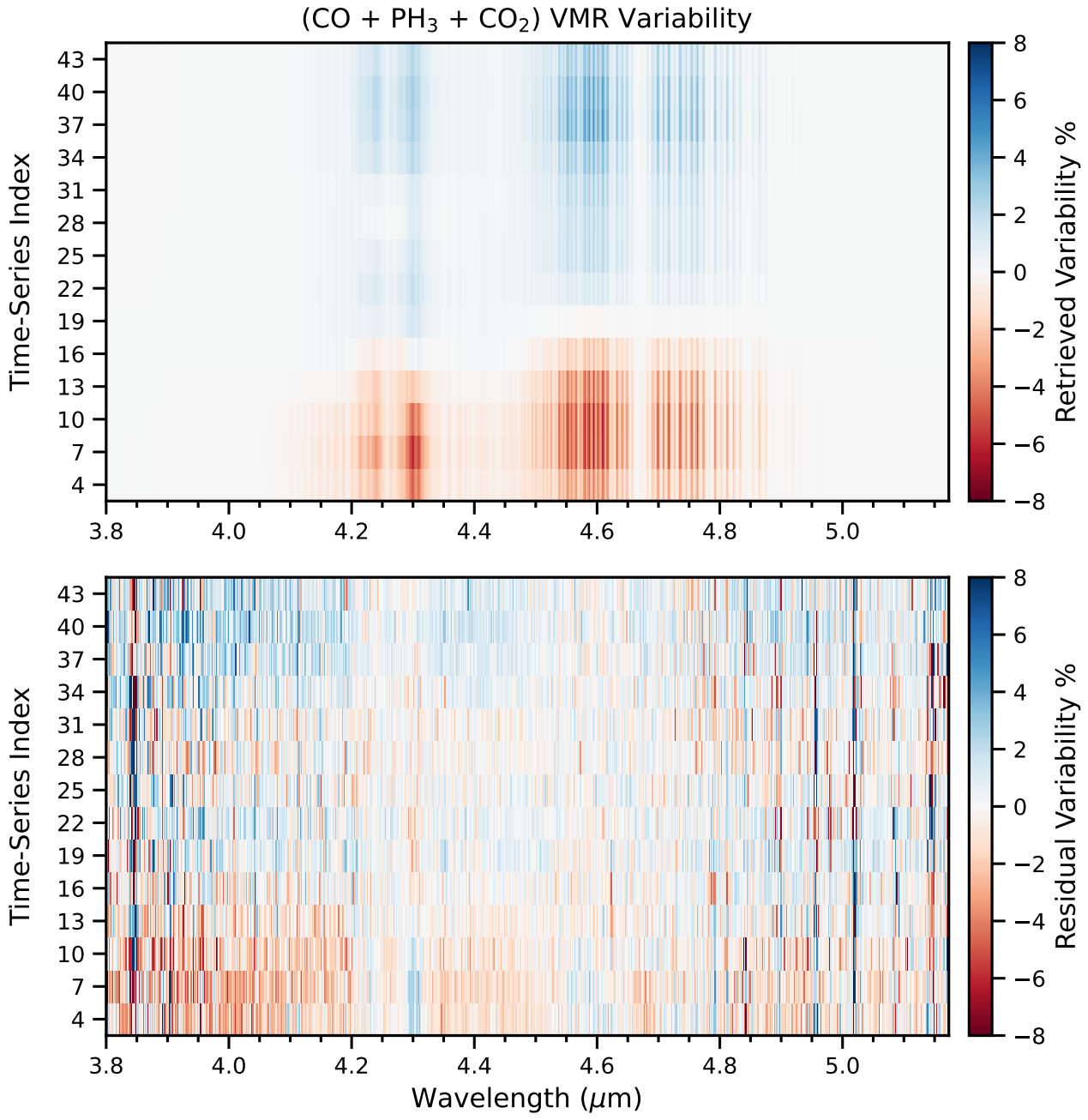

**Figure 8.** Spectral variability captured by the ($CO + CO_2 + PH_3$) VMR variability model. The top panel shows the retrieved spectra normalized to the overall mean model spectrum, illustrating the variability reproduced when CO, $PH_3$, and $CO_2$ VMRs are all allowed to vary. The bottom panel shows the residual variability percent between the observed and modeled spectra, with both normalized by their respective mean spectra. This model scenario yields a weighted rms residual variability of 1.83%.

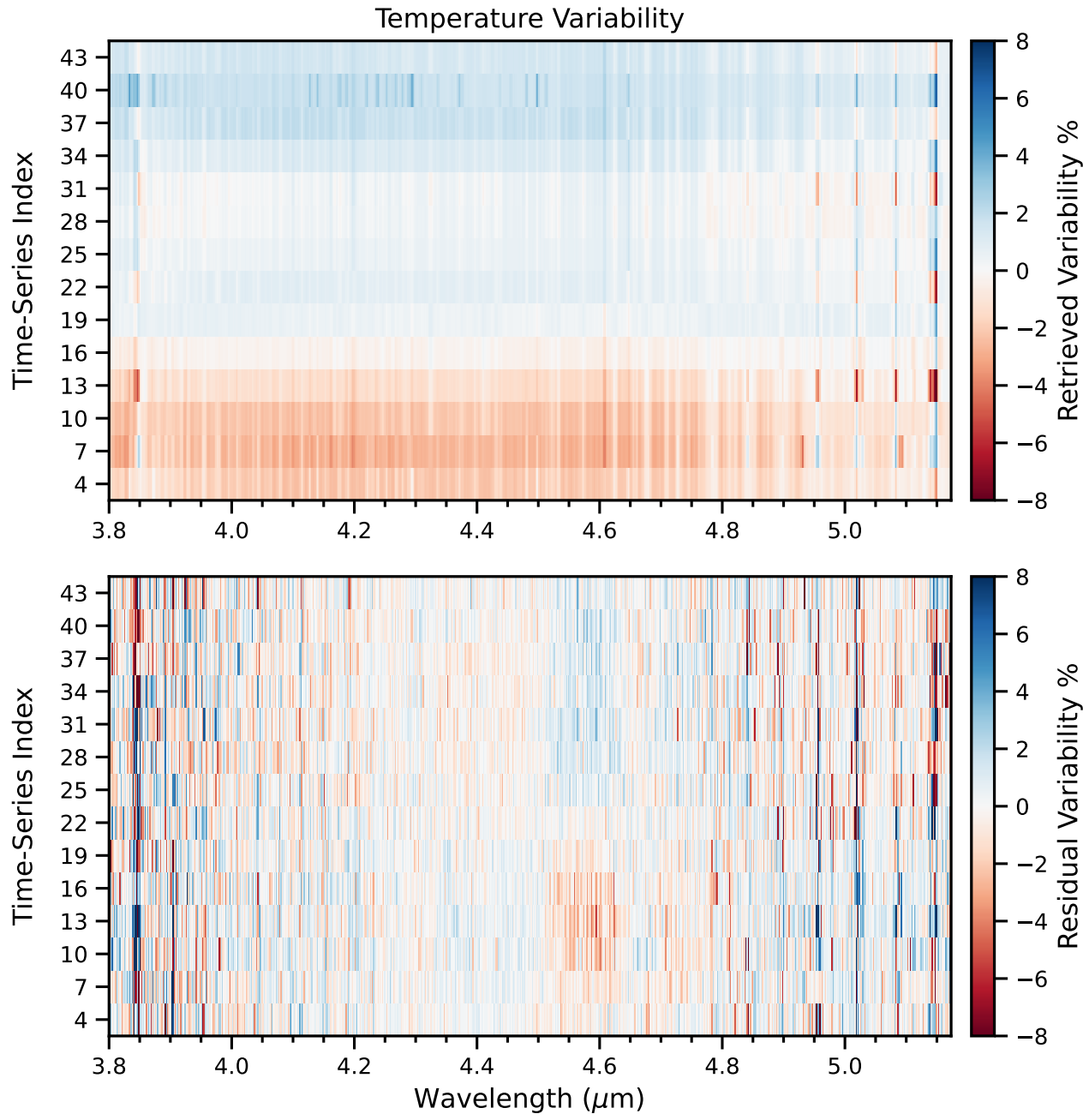

**Figure 9.** Spectral variability captured by the temperature variability model. The top shows the retrieved spectra normalized to the mean model spectrum, illustrating the variability reproduced when only the thermal profile is allowed to vary. The bottom panel shows the residual variability percent between the observed and modeled spectra, with both normalized by their respective mean spectra. This model scenario yields a weighted rms residual variability of 1.75%.

account for the overlapping $CO_2$ and $PH_3$ absorption in this region.

This improvement is also reflected quantitatively as the weighted rms residual variability decreases further to 1.83%. Nevertheless, this value still remains above the noise floor, and significant residuals persist, particularly shortward of 4.0 $\mu$m and toward the long-wavelength end of the bandpass. Thus, even when the VMRs of CO, $PH_3$, and $CO_2$ are all allowed to vary, the model does not reproduce the full amplitude and wavelength dependence of the observed variability.

Taken together, the VMR-only models indicate that abundance changes can explain part of the observed spectral modulation, especially within the major molecular bands, but do not by themselves account for the full observed variability pattern of WISE 0855−07.

### *4.3. Temperature-only Variability*

We next test the complementary hypothesis that the observed variability is primarily driven by changes in the thermal structure. In this case, the molecular VMRs and all remaining atmospheric parameters are held fixed to their fiducial median values, and only the thermal profile is allowed to vary from epoch to epoch.

To test whether thermal perturbations alone can drive the observed variability, we performed 14 independent retrievals in which only the thermal profile was allowed to vary. Specifically, we retrieved the hyperparameter $\gamma$ together with the temperatures at the nine pressure knots for each spectrum, while fixing all other parameters to their fiducial median values.

Figure 9 shows the variability reproduced by this temperature-only model. The top panel presents the retrieved spectra divided by the overall mean model spectrum, and demonstrates that variations in the thermal structure alone reproduce a large fraction of the observed wavelength-dependent variability. In contrast to the VMR-only scenarios, the temperature-only model captures substantial variability across most of the full 3.8–5.2 $\mu$m interval.

This improvement is also evident in the bottom panel, where the residual variability percent is markedly reduced relative to the VMR-only models. Quantitatively, the weighted rms residual variability decreases to 1.75%, substantially lower than in the VMR-only cases and much closer to the 1.68% noise floor. This indicates that changes in the thermal profile provide a substantially better description of the data than VMR variations alone.

Overall, these results show that temperature variability is a strong candidate for the dominant driver of the observed variability in WISE 0855−07. Molecular VMR changes may still contribute, but the temperature-only model already captures most of the observed variability.

### *4.4. Temperature + VMR Variability*

Finally, we consider models in which both the thermal structure and selected molecular VMRs are allowed to vary. These scenarios test whether adding VMR variability to the temperature-driven models improves the fit within specific molecular bands and provides a more complete description of the observed wavelength-dependent variability.

#### *4.4.1. Temperature + CO VMR Variability*

We next considered a model in which both the thermal profile and the CO VMR were allowed to vary across the 14 retrievals, while all other atmospheric parameters were fixed to their fiducial median values. This scenario tests whether adding CO abundance variability to the temperature-only model improves the fit in the CO-dominated portion of the spectrum.

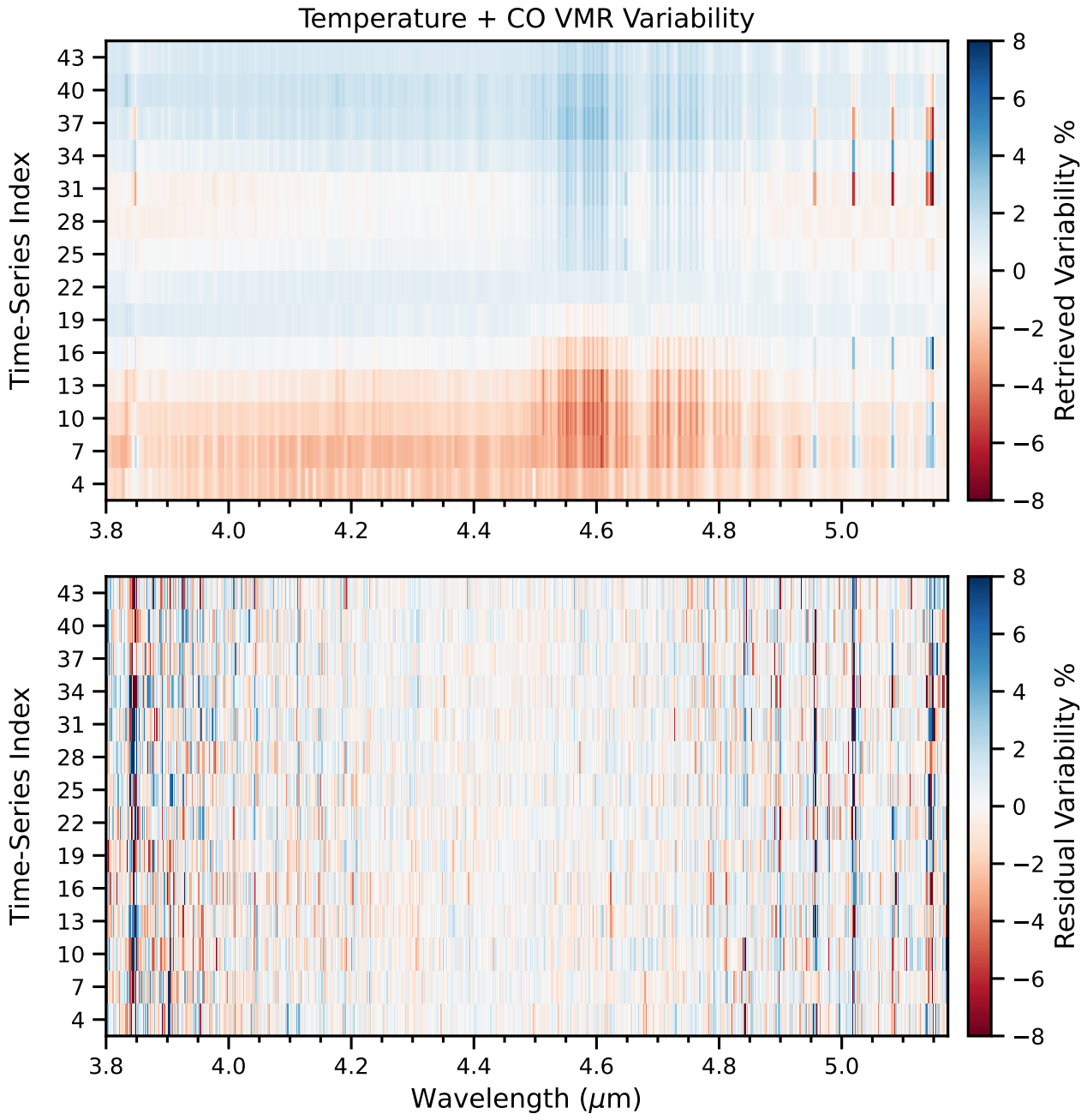

**Figure 10.** Spectral variability captured by the combined temperature + CO VMR variability model. The top panel shows the retrieved spectra normalized to the mean model spectrum, illustrating the variability reproduced when both the thermal profile and the CO VMR are allowed to vary. The bottom panel shows the residual variability percent between the observed and modeled spectra, with both normalized by their respective mean spectra. This model scenario yields a weighted rms residual variability of 1.68%.

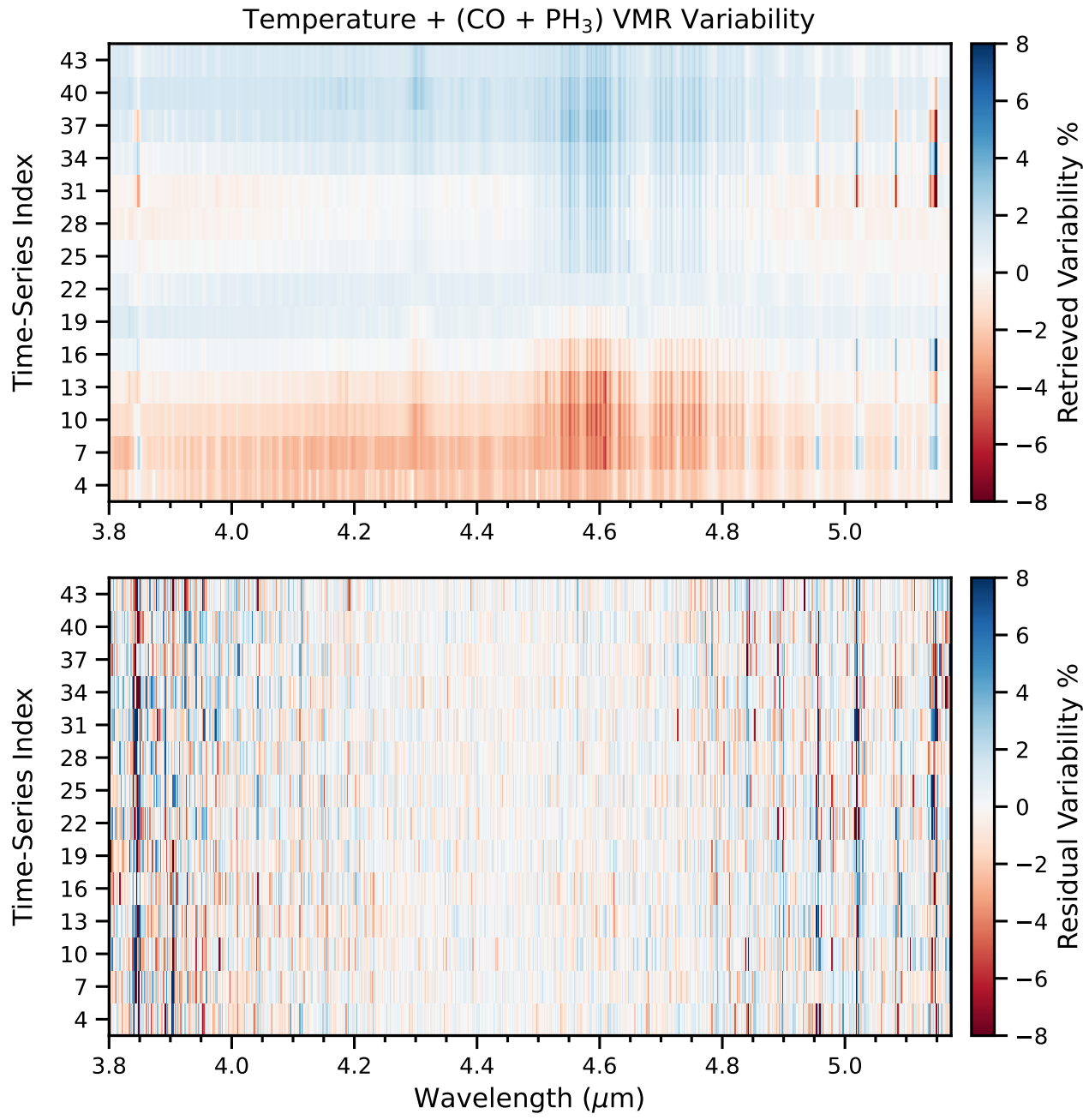

**Figure 11.** Spectral variability captured by the combined temperature + (CO + $PH_3$) VMR variability model. The top panel shows the retrieved spectra normalized to the mean model spectrum, illustrating the variability reproduced when the thermal profile and the CO and $PH_3$ VMRs are allowed to vary. The bottom panel shows the residual variability percent between the observed and modeled spectra, with both normalized by their respective mean spectra. This model scenario yields a weighted rms residual variability of 1.68%.

Figure 10 shows the resulting variability map. The top panel demonstrates that the combined temperature + CO model reproduces the overall wavelength-dependent variability pattern across most of the 3.8–5.2 $\mu$m range. The bottom panel shows that the residual variability percent remains small over most wavelengths, indicating an improved match relative to both the VMR-only and temperature-only models.

The most evident improvement relative to the temperature-only case occurs across the CO $v = 1$–0 fundamental band at 4.50–4.90 $\mu$m, where allowing the CO abundance to vary further reduces the residual structure. This is also reflected in the weighted rms residual variability, which decreases to 1.68%, reaching the estimated noise floor. Residuals that persist at the longest wavelengths are comparatively small and may reflect lower signal-to-noise and/or calibration systematics near the edge of the spectral range.

This model therefore supports a picture in which temperature perturbations provide the primary explanation for the observed variability, while CO VMR variations make a secondary but measurable contribution within the strongest CO absorption band.

#### 4.4.2. Temperature + (CO + $PH_3$) VMR Variability

We also explored a model in which the thermal profile, together with the VMRs of CO and $PH_3$, were allowed to vary across epochs. As before, all remaining atmospheric parameters were fixed to their fiducial median values. This scenario tests whether adding $PH_3$ variability, in addition to CO variability, improves the fit in the 4.1–4.5 $\mu$m interval while retaining the strong performance of the temperature-driven model.

Figure 11 presents the results. The normalized model spectra in the top panel reproduce the observed wavelength-dependent variability across nearly the full 3.8–5.2 $\mu$m range, while the bottom panel shows low residual variability percent over most wavelengths. Relative to the temperature + CO model, the inclusion of $PH_3$ variability marginally improves the fit in the 4.1–4.5 $\mu$m region, including the portion of the spectrum where the $PH_3$ and $CO_2$ opacities overlap near 4.2–4.3 $\mu$m.

This qualitative improvement occurs without a further reduction in the global weighted rms residual variability, which remains 1.68%, equal to the estimated noise floor and identical to the temperature + CO VMR variability model. Thus, while allowing $PH_3$ to vary marginally improves the phenomenological agreement in specific spectral regions, it does not produce an additional statistically measurable improvement in the overall fit according to this metric.

These results again point to temperature fluctuations as the dominant source of the observed variability, with secondary contributions from composition changes in disequilibrium species.

#### 4.4.3. Temperature + (CO + $PH_3$ + $CO_2$) VMR Variability

Finally, we considered the most flexible model in our grid, in which the thermal profile and the VMRs of CO, $PH_3$, and $CO_2$ were all allowed to vary across the 14 retrievals, while the remaining parameters were fixed to their fiducial median values. This model tests whether the combined effects of temperature perturbations and abundance changes in all three species can account for the full observed variability pattern.

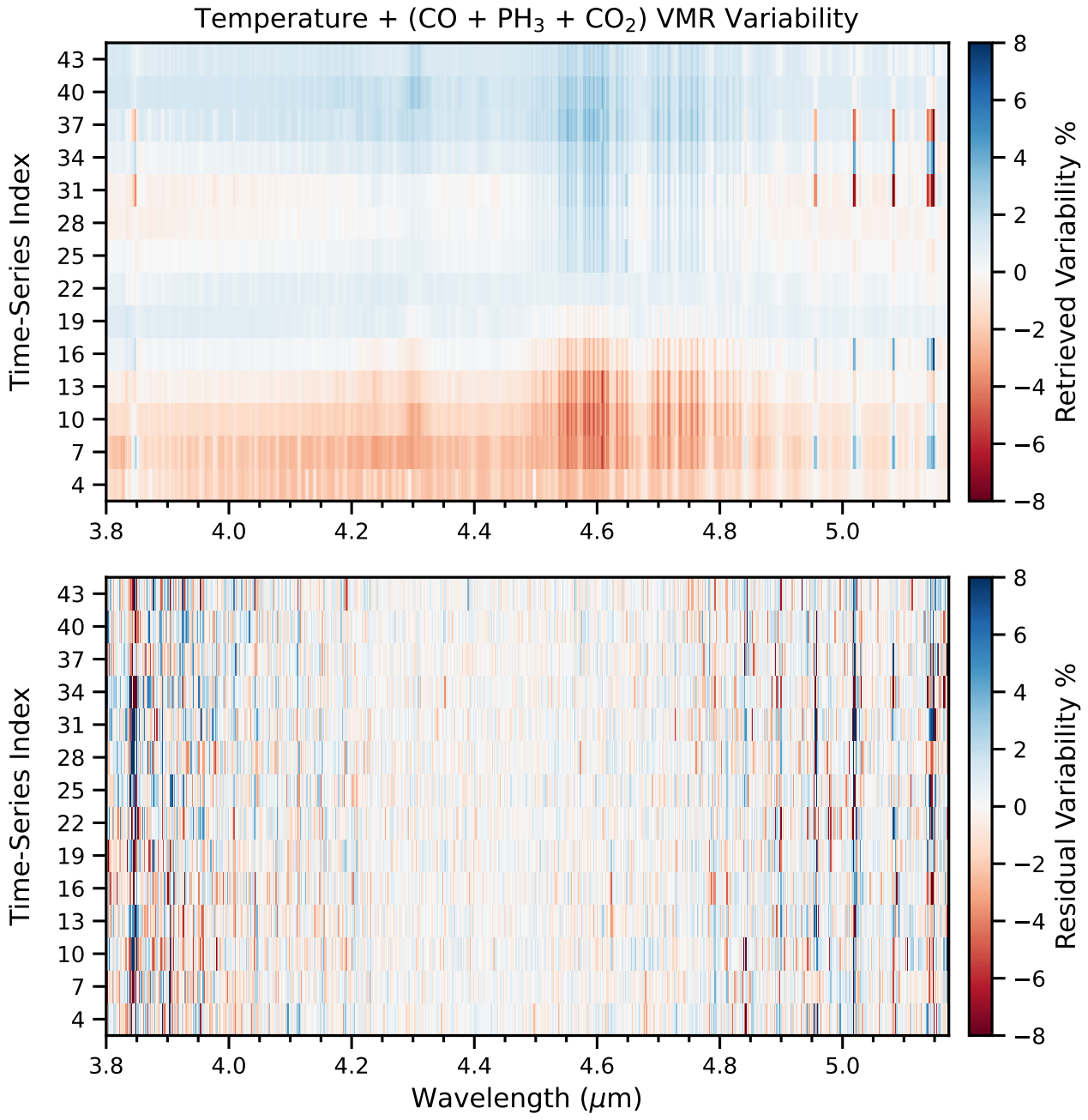

**Figure 12.** Spectral variability captured by the temperature + (CO + $PH_3$ + $CO_2$) VMR variability model. The top panel shows the retrieved spectra normalized to the mean model spectrum, illustrating the variability reproduced when the thermal profile and the CO, $PH_3$, and $CO_2$ VMRs are all allowed to vary. The bottom panel shows the residual variability percent between the observed and modeled spectra, with both normalized by their respective mean spectra. This model scenario yields a weighted rms residual variability of 1.68%.

Figure 12 shows the corresponding results. The top panel indicates that this model reproduces the observed variability structure across the major molecular features, including the CO fundamental band at 4.50–4.90 $\mu$m, the $PH_3$-sensitive 4.1–4.5 $\mu$m region, and the $CO_2$-affected overlap region near 4.2–4.3 $\mu$m. The bottom panel shows that the residual variability percent remains low across most of the spectral range.

Compared with the other models, this scenario provides the most complete phenomenological description of the observed wavelength-dependent variability. However, the weighted rms residual variability still remains 1.68%, identical to both the temperature + CO VMR and temperature + (CO + $PH_3$) VMR variability models and therefore also at the noise floor. In other words, once temperature and CO VMR variability are included, allowing $PH_3$ and $CO_2$ to vary improves the detailed agreement within specific molecular bands, but does not yield an additional statistically measurable improvement in the global fit.

We therefore conclude that the observed spectral evolution of WISE 0855−07 is best explained by variability in the thermal structure, with additional contributions from VMR changes in disequilibrium species, particularly from CO.

### 4.5. Comparison of Retrieval Variability Model Scenarios

The weighted rms residual variability values introduced throughout the preceding subsections provide a direct quantitative ranking of the seven tested variability scenarios. This statistic compresses the full wavelength–time residual variability map into a single measure of the typical unexplained variability remaining after subtraction of a given model, expressed in percent and weighted by the measurement uncertainties. Smaller values therefore indicate that a model captures a larger fraction of the observed variability, while values approaching the 1.68% noise floor imply that the remaining residual structure is statistically consistent with measurement noise.

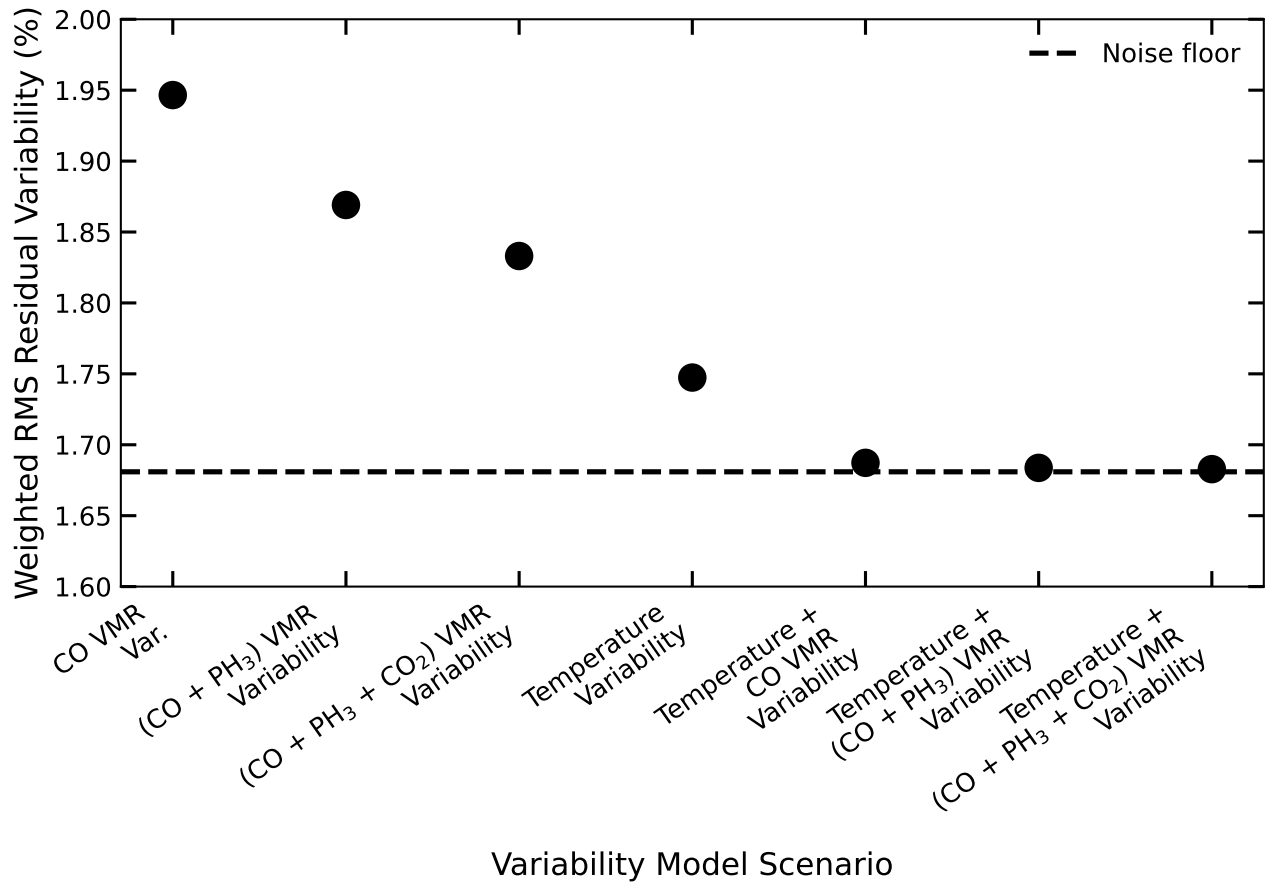

**Figure 13.** Comparison of the weighted rms residual variability percent for the seven variability model scenarios considered for WISE 0855−07. Smaller values indicate that a model leaves less unexplained variability after accounting for the spectral uncertainties. The dashed horizontal line marks the estimated noise floor.

Figure 13 summarizes this ranking. The VMR-only models yield the largest weighted rms residual variability values, decreasing from 1.95% for the CO-only VMR variability model to 1.87% for the CO + $PH_3$ VMR variability model, and to 1.83% when $CO_2$ is also allowed to vary. These results show that molecular VMR changes alone improve the fit within specific spectral bands, but do not account for the dominant wavelength-dependent variability pattern.

A much larger improvement is obtained when only the thermal profile is allowed to vary. The temperature-only model yields a weighted rms residual variability of 1.75%, substantially lower than the chemistry-only cases and much closer to the noise floor. This demonstrates that temperature perturbations capture most of the observed variability across the full spectral range.

The lowest residual variability values are obtained for the combined temperature + VMR models. The temperature + CO, temperature + (CO + $PH_3$) and temperature + (CO + $PH_3$ + $CO_2$) models all yield 1.68%, reaching the estimated noise floor. This indicates that once temperature variability is included, allowing CO variability is sufficient to reduce the unexplained residual variability to the level expected from the propagated measurement uncertainties. Allowing $PH_3$ and $CO_2$ to vary further improves the phenomenological agreement within specific molecular-band regions, but does not provide an additional statistically measurable improvement in the global weighted rms metric.

Taken together, the ordering of the models from largest to smallest weighted rms residual variability implies a clear physical hierarchy among the variability drivers. Temperature variations account for most of the observed wavelength-dependent temporal modulation, while VMR variations in disequilibrium species, particularly CO, provide secondary refinements that improve the fit in specific spectral intervals.

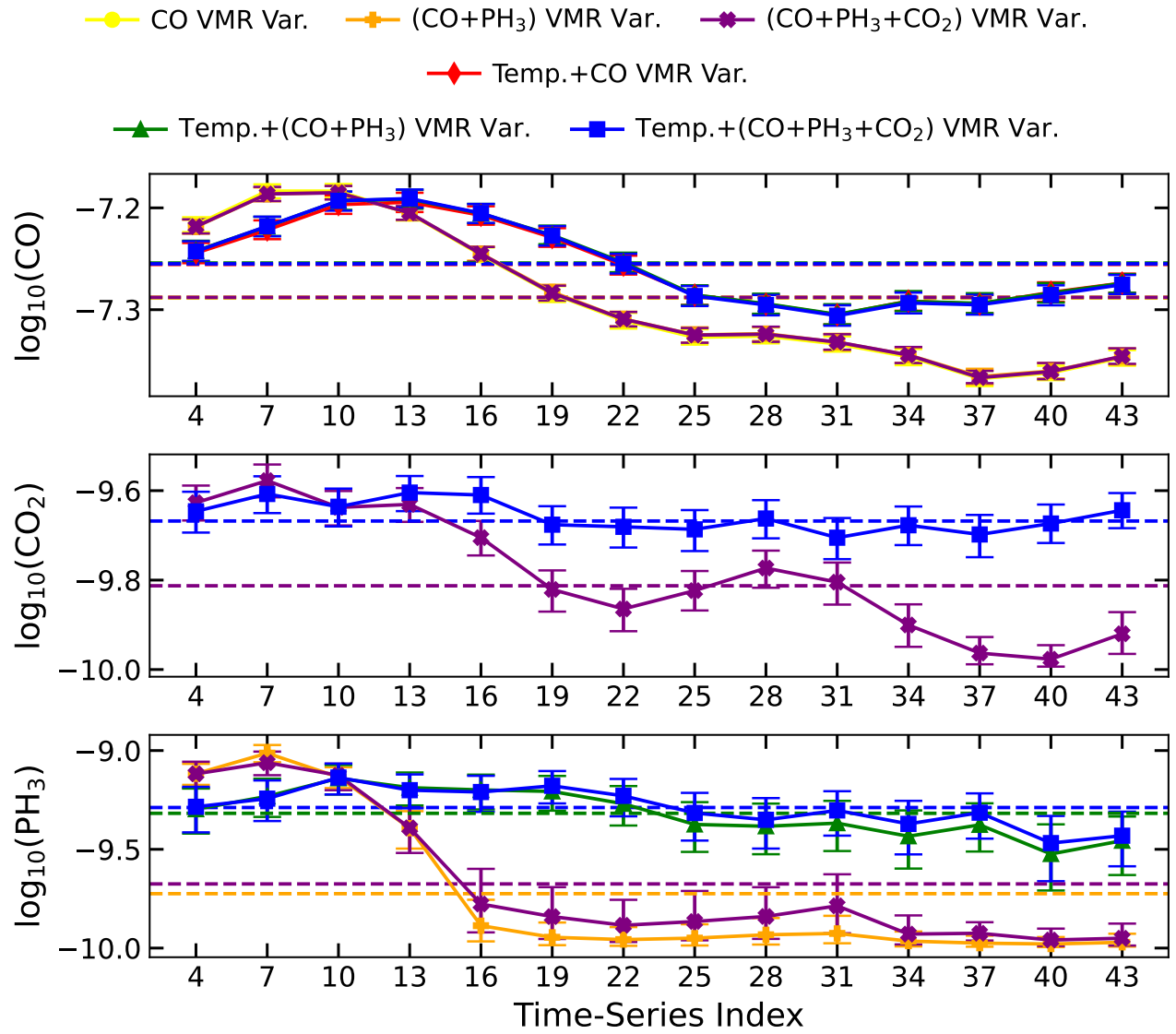

**Figure 14.** Retrieved VMR from each variability model that allows the VMR to vary as a function of time-series index. The panels show the retrieved median VMR for CO, $PH_3$, and $CO_2$ with the error bars representing the 1$\sigma$ credible intervals for each variability model.

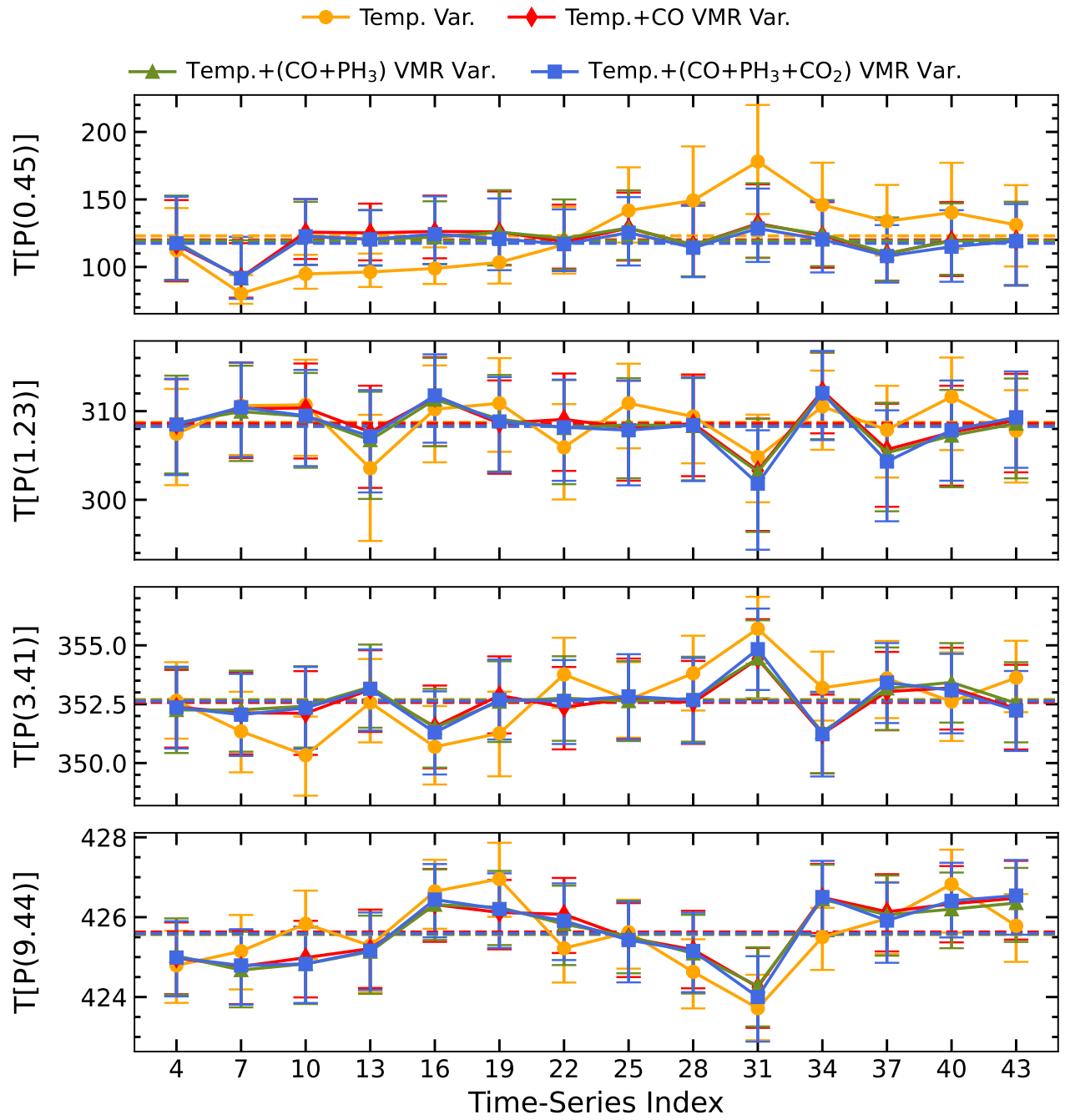

**Figure 15.** Retrieved temperature structure in kelvin from each variability model that allows the thermal profile to vary as a function of time-series index. The panels show the retrieved median temperature for four pressure knots with the error bars representing the 1$\sigma$ credible intervals for each variability model.

## 5. Discussion

We have so far discussed how individual retrieval models capture the observed variability. In this section, we discuss how the VMR for CO, $PH_3$, and $CO_2$, and the temperature structure changes temporally for the part of the atmosphere probed by our time-series dataset. We also ran another retrieval using the mean spectrum and compared the results with the previous work done by M. J. Rowland et al. (2024), who also used the mean spectrum from the same dataset, and we briefly discuss the future role of a multicolumn retrieval model to study spectral variability.

### 5.1. VMR−Temperature Variability Comparison

As mentioned in Section 4, we tested a sequence of retrieval models with increasing complexity to assess the physical drivers of the observed variability in WISE 0855−54. Figures 14 and 15 summarize the resulting molecular VMR variations in CO, $CO_2$, and $PH_3$ and temperature variations at four pressure knots [P(0.45), P(1.23), P(3.41), and P(9.44)], which covers the atmospheric pressure region that is probed by our dataset.

Across all retrieval configurations, the CO VMR exhibits significant variability over the entire time-series spectral indices. Both the peak-to-peak amplitude and temporal morphology of the CO VMR variation remain within 1$\sigma$ whether (1) only CO is permitted to vary or (2) additional molecular species ($PH_3$, $CO_2$) are allowed to vary simultaneously. The consistency of CO VMR variation in VMR-only variation and temperature + VMR variation, respectively, indicates that CO is a dominant chemical driver of the observed time-dependent flux modulations. Even in the temperature + VMR variation models, the CO VMR variation retains a comparable amplitude to that of VMR-only variation models, confirming that the retrieval does not artificially attribute spectral structure to temperature when CO variability is permitted.

In contrast to CO VMR variation, $PH_3$ VMR variations remain small in amplitude and largely consistent (within 1$\sigma$) across all VMR-only and temperature + VMR models, respectively. Allowing $CO_2$ to vary alters the median $PH_3$ VMR slightly but does not introduce significant temporal structure.

The temperature-only and temperature + VMR models (Figure 15) show that retrieved temperatures at P(1.23), P(3.41), and P(9.44) pressure knots vary only by a few kelvin, well within the 1$\sigma$ uncertainties. The small changes in the temperature for the aforementioned pressure knots indicate that the observed variability due to temperature fluctuations is only slightly influenced by the deep atmosphere. However, the pressure knot at P(0.45) bar shows a large variation in temperature, with a delta of ∼100 K, driving the observed variability strongly.

### 5.2. Mean Spectrum Retrieval Comparison

M. J. Rowland et al. (2024) presented a retrieval analysis of the mean time-series spectrum of WISE 0855 using the `CHIMERA` retrieval framework (M. R. Line et al. 2014). Our retrieval setup is broadly similar, but differs in several important respects:

1. We exclude $^{13}$CO from our retrieval model, following M. J. Rowland et al. (2024), who showed that $^{13}$CO is both unconstrained and negligible in abundance.
2. We adopt a higher and variable linelist with an equivalent resolving power of 40,000−100,000, compared to the lower variable equivalent resolving power of 18,000−35,000 used by M. J. Rowland et al. (2024). The higher resolution provides improved fidelity in modeling pressure-broadened line shapes.

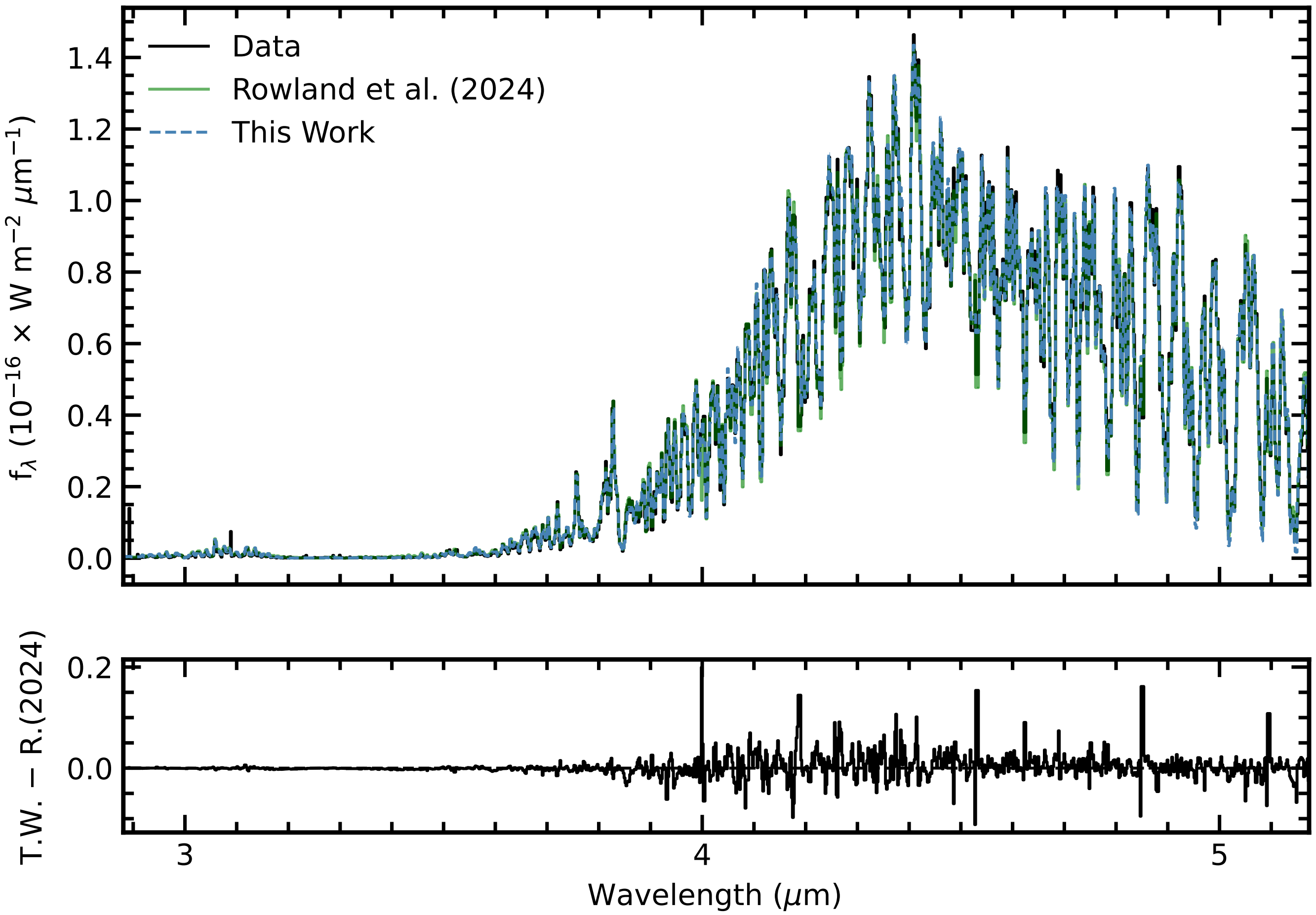

**Figure 16.** Comparison of the retrieved median spectra from the mean observed spectrum in this work and in M. J. Rowland et al. (2024). The top panel shows the median retrieved spectrum from this work in blue and that from M. J. Rowland et al. (2024) in green with the data and the corresponding $1\sigma$ uncertainties in black and gray, respectively. The bottom panel shows the wavelength-dependent difference between the two median model spectra.

3. We parameterize the thermal profile using nine pressure knots across the pressure range from $10^{-3}$ to $10^{2.3}$ bar, with knot locations informed by the contribution functions in order to provide finer sampling in the regions most strongly probed by the spectrum. By contrast, M. J. Rowland et al. (2024) used 18 equally spaced points spanning $10^{-4.3}$ to $10^{2.35}$ bar.

Figure 16 compares our retrieval of the mean observed spectrum to that of M. J. Rowland et al. (2024). The top panel shows the retrieved median spectra from both works, while the bottom panel shows their difference. Although the two models are broadly consistent, noticeable differences are present in several spectral regions, particularly near 4.0, 4.1–4.2, 4.5–4.6, and 4.8–4.9 $\mu$m. These discrepancies likely reflect the modeling differences listed above, especially the treatment of linelist resolution and thermal profile parameterization. An additional source of difference may arise from the data reduction itself: our analysis uses spectra reprocessed with a newer version of the JWST pipeline, whereas M. J. Rowland et al. (2024) analyzed an earlier reduction.

These differences in model setup and data reduction can translate into differences in the retrieved temperature structure. Figure 17 compares the thermal profiles inferred from the mean spectrum in this work and by M. J. Rowland et al. (2024). The left panel shows that the two median profiles agree well throughout much of the pressure range most strongly probed by the data, as indicated by the mean contribution function. However, systematic offsets are present at both lower and higher pressures. The right panel shows that, relative to M. J. Rowland et al. (2024), our retrieved profile is generally cooler in the upper atmosphere, becomes modestly warmer at intermediate pressures near the peak of the contribution function, and then becomes cooler again at deeper pressures. This behavior suggests that, although both retrievals infer a broadly similar temperature structure over the pressure range to which the spectrum is most sensitive, differences in the thermal profile parameterization, model assumptions, and data reduction can lead to nonnegligible changes in the detailed shape of the retrieved thermal profile, particularly outside the region of strongest observational coverage.

These differences in the thermal structure naturally affect the inferred atmospheric VMRs. Figure 18 compares the posterior distributions for the VMRs of common gases and the surface gravity ($\log_{10}(g)$, calculated as $g = GM/R^2$ in cm/s$^2$ from the retrieved mass ($M$) and radius ($R$), where $G$ is the gravitational constant, $6.67430 \times 10^{-11}$ m$^3$ kg$^{-1}$ s$^{-2}$; E. E. Mamajek et al. 2015) between this work and M. J. Rowland et al. (2024). For most species, the posterior distributions overlap significantly, indicating broad consistency in the inferred atmospheric composition. However, there are notable shifts in the preferred VMRs of some molecules. In particular, our retrieval favors marginally higher VMRs of $H_2O$, $CH_4$, CO, $NH_3$, and $H_2S$, and lower VMRs of $CO_2$ and $CH_3D$ relative to M. J. Rowland et al. (2024).

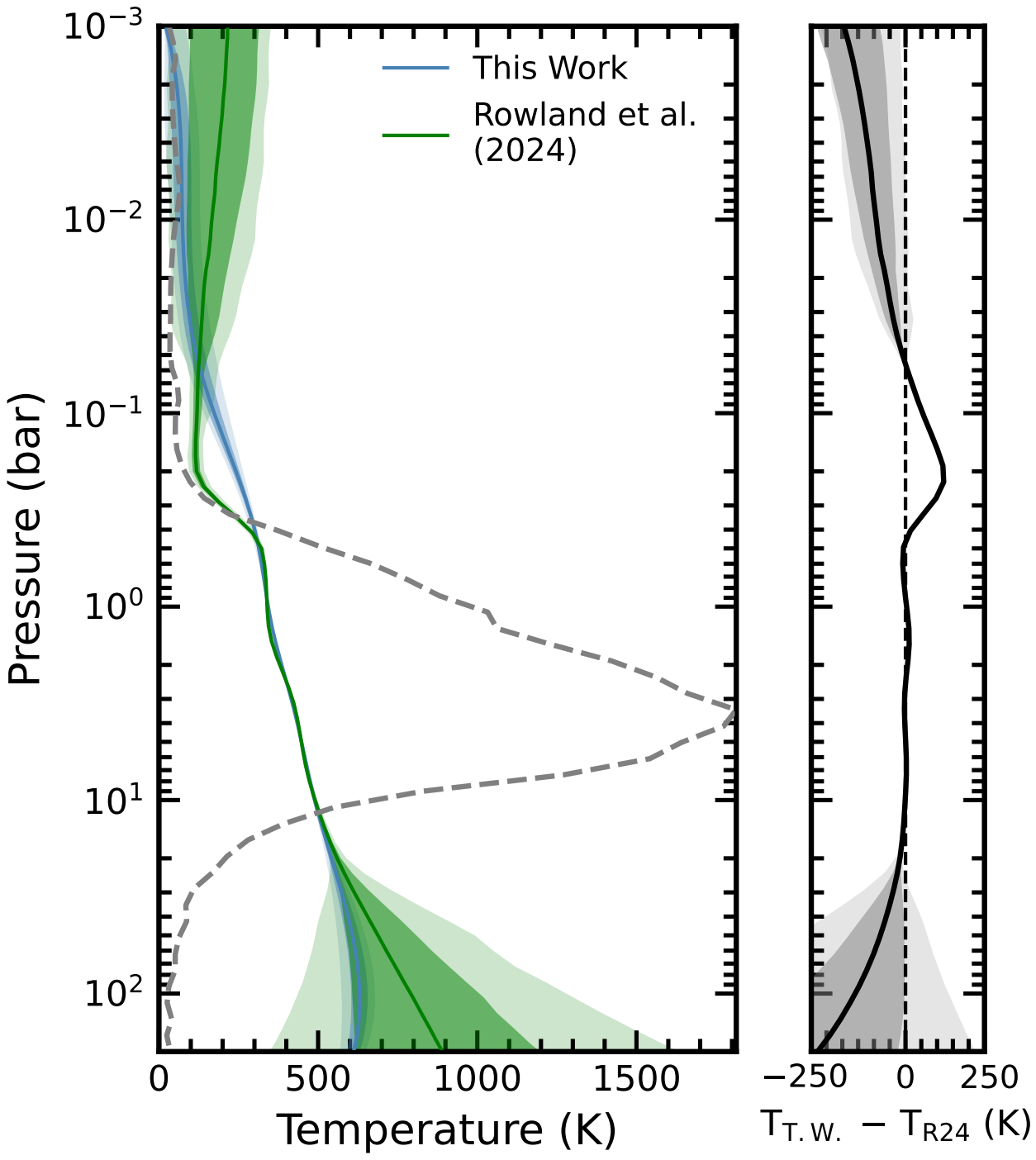

**Figure 17.** Comparison of the retrieved thermal profiles from the mean observed spectrum in this work and in M. J. Rowland et al. (2024). The left panel shows the median retrieved thermal profile from this work in blue and that from M. J. Rowland et al. (2024) in green. The gray dashed curve shows the mean contribution function and indicates the pressure levels that contribute most strongly to the observed flux. The right panel shows the temperature difference between the two retrieved thermal profiles, defined as $T_{\rm ThisWork} - T_{\rm Rowlandetal.\ (2024)}$.

Overall, this comparison highlights that retrieval results remain broadly robust at the qualitative level, but can still show meaningful differences in the detailed thermal structure and molecular VMRs as a result of model assumptions, opacity treatment, and data reduction choices.

This comparison highlights the sensitivity of retrieval results based on the model assumptions and data quality.

### 5.3. Future Work: Multicolumn Retrieval Model

The fiducial, cloud-free retrieval framework adopted here provides a good description of the NIRSpec/G395M time-series spectra, and the residual structure is generally small across the 3.8–5.2 $\mu$m range. While our 1D retrieval model captures both the primary and secondary observed variability from thermal and compositional variations, respectively, we would like to emphasize that a single, homogeneous atmospheric column is a rudimentary atmospheric model. Variability in brown dwarfs is widely interpreted as evidence for atmospheric heterogeneity, arising from spatial structure in temperature, composition, and cloud opacity (D. Apai et al. 2013; J. M. Vos et al. 2023; B. A. Biller et al. 2024; A. M. McCarthy et al. 2025). In this broader context, an attractive extension of the present analysis is the use of multicolumn retrieval models, in which the visible atmosphere is represented by a small number of distinct vertical columns (most commonly two) that capture different atmospheric states. Conceptually, such columns can be interpreted as "background" and "perturbed" regions, or as representative patches with different thermal structures and/or cloud properties. Recent JWST-era work, like Z. Zhang et al. (2025), has demonstrated that multicolumn approaches can provide a physically transparent way to model atmospheric heterogeneity while remaining computationally tractable for retrieval analyses (J. M. Vos et al. 2023).

Applied to the WISE 0855–07 time-series, a multicolumn framework would offer a complementary way to interpret the variability patterns recovered by our epoch-by-epoch retrievals. Rather than treating each epoch as a fully independent atmospheric state, a multicolumn model would aim to identify a small set of persistent atmospheric components whose relative prominence changes with time. This can be particularly useful when the data show coherent, repeatable variability features as it encourages an interpretation in which the atmosphere is composed of a limited number of characteristic regions, such as warmer versus cooler areas, or clearer versus more opaque patches, whose changing visibility produces the observed modulation. In practice, this framing can sharpen physical interpretation by separating "what kinds of atmospheric states exist" from "how their apparent contributions evolve across the time series" (E. Nasedkin et al. 2025; F. Wang et al. 2026).

A multicolumn model is also naturally aligned with the cloud motivation discussed in the literature for ultracold atmospheres. In the temperature regime of WISE 0855−071, models predict that condensate clouds, including water-ice clouds, may form at high altitudes and plausibly introduce spatial inhomogeneity in the observable photosphere (C. V. Morley et al. 2012, 2014b). Observational studies in the 4.5–5.2 $\mu$m region have highlighted strong water absorption and demonstrated that the overall spectral morphology is consistent with a cold, Jupiter-like atmosphere in which clouds play an important role in shaping the emergent flux (A. J. Skemer et al. 2016). In this setting, a future multicolumn retrieval could be used to test cloud-related hypotheses in a way that is more flexible than a globally cloud-free versus globally cloudy dichotomy. For example, one atmospheric column could represent a comparatively clearer region while another corresponds to a more opaque, cloudier region, analogous to heterogeneous or patchy cloud models that have been successfully applied to warmer brown dwarfs (e.g., J. Radigan et al. 2012; D. Apai et al. 2013; E. Buenzli et al. 2014; J. M. Vos et al. 2023). Alternatively, one column could encode a warmer, deeper-emitting state while the other represents a cooler configuration in which the photosphere is shifted to higher altitudes. In this framework, condensate clouds contribute primarily by generating contrast between regions with different thermal structures and opacity profiles, providing a physically motivated mechanism for rotationally modulated variability in ultracool atmospheres.

The most powerful role for multicolumn retrievals of WISE 0855–07 will likely emerge if the time-series analysis is extended to include mid-infrared spectroscopy that probes cooler, higher layers where condensate clouds exert stronger influence. JWST/MIRI spectroscopy provides precisely this leverage and is expected to be critical for breaking degeneracies between cloud opacity and thermal structure. In that regime, a multicolumn retrieval could become a stringent test of whether the variability is dominated by temperature contrasts, cloud opacity contrasts, or a coupled combination of the two, while maintaining consistency across epochs through a small number of physically interpretable atmospheric components.

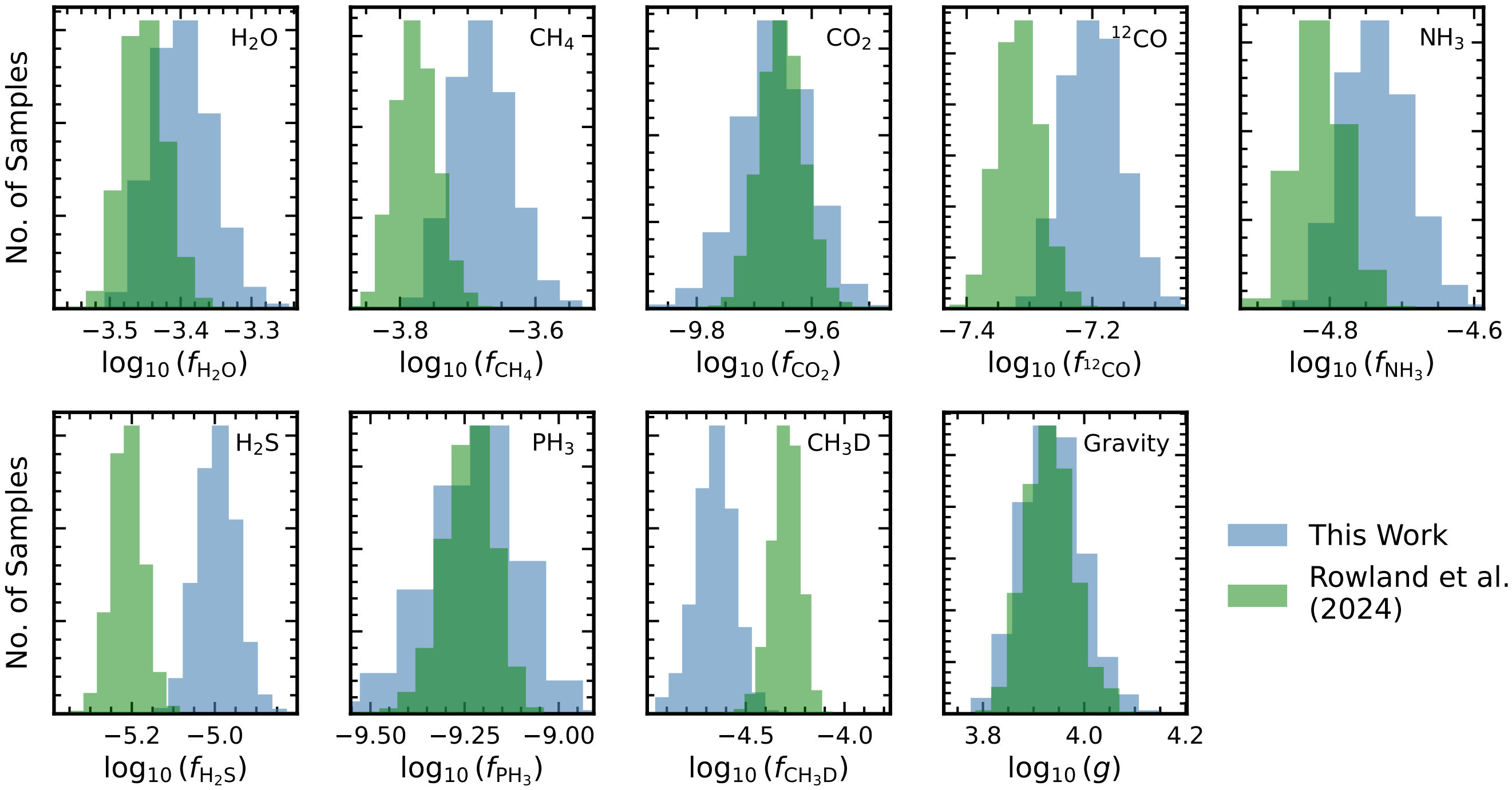

**Figure 18.** Comparison of the posterior distributions from the mean spectrum retrieval in this work and in M. J. Rowland et al. (2024). The distributions shown correspond to the retrieved PDFs for $H_2O$, $CH_4$, CO, $CO_2$, $NH_3$, $H_2S$, $PH_3$, and $CH_3D$, as well as surface gravity, where $\log_{10}(g)$ is calculated from the retrieved mass and radius. Blue denotes this work and green denotes M. J. Rowland et al. (2024).

Despite its physical appeal, implementing a full multicolumn retrieval analysis is beyond the scope of the present work. Such a model would substantially increase both the dimensionality of the parameter space and the computational cost of forward model evaluations, particularly in the context of time-resolved JWST datasets where multiple epochs must be modeled. Moreover, developing a robust multicolumn framework requires careful consideration of parameter degeneracies and column parameterizations for linking atmospheric components across the time-series dataset. This challenge is further compounded by the need for well-constrained rotational phase information: robustly associating atmospheric structures across epochs generally requires time series observations spanning multiple rotation periods and an accurately determined rotation period. In the absence of such constraints, linking atmospheric components becomes inherently ambiguous. The primary goal of this study is instead to establish a description of the variability in WISE 0855−07 using a well-controlled 1D retrieval framework and to characterize the dominant thermal and compositional drivers of the observed modulation. Building on these results, a dedicated multicolumn retrieval analysis would represent a natural and promising avenue for future work aimed at probing atmospheric heterogeneity in ultracold brown dwarfs.

## 6. Conclusions

We have presented the first time-resolved atmospheric retrieval analysis of the coldest known brown dwarf, WISE 0855−07, using every third spectrum from the dataset of 44 JWST/NIRSpec G395M spectra spanning 3.8–5.2 $\mu$m observed over an ~11 hr baseline. By applying the `Brewster` retrieval framework to individual epochs and systematically testing a hierarchy of variability model scenarios, we directly constrain variations in the thermal structure and key molecular VMRs on observable timescales in an ultracool, planet-like atmosphere.

Our fiducial retrieval of the first spectrum from the dataset demonstrates that a cloud-free atmosphere with variable linelist resolving power ($R$ = 40,000–100,000) provides an adequate fit to the data. To isolate the physical drivers of variability, we tested model scenarios including VMR-only variability, temperature-only variability, and joint temperature + VMR variability models. We find the following:

1. VMR-only variability models (CO; CO + $PH_3$; CO + $PH_3$ + $CO_2$) reproduce variability within specific molecular bands, particularly the CO $v$ = 1–0 fundamental band (4.5–4.9 $\mu$m), the $PH_3\nu_4$ band, and the $CO_2\nu_3$ band, but fail to capture the full wavelength-dependent modulation across 3.8–5.2 $\mu$m.
2. Temperature-only variability reproduces most of the observed variability across the full spectral window and significantly reduces the residual structure relative to the VMR-only models. In this model, the retrieved temperature variations are largest at the upper pressure knot near $P \approx 0.45$ bar, reaching nearly ~100 K, while variations at deeper pressures of 3.41 and 9.44 bar remain much smaller, typically only a few kelvin (see Figure 15).
3. Combined temperature + VMR variability models provide the best overall match to the data. When molecular abundance variability is included together with temperature, the required temperature perturbations become smaller, particularly at $P \approx 0.45$ bar, where the maximum variation decreases from ~100 K in the temperature-only case to roughly ~40 K for the temperature + VMR models. The temperature + CO, temperature + (CO + $PH_3$), and temperature + (CO + $PH_3$ + $CO_2$) models capture both the broad continuum-level modulation and structured variability within molecular bands, leaving only minor residuals. This suggests that the observed variability is best explained by a combination of thermal and compositional changes, rather than by temperature fluctuations alone.

These qualitative trends are quantitatively supported by the noise-weighted rms residual variability comparison (Section 3.7). The VMR-only models yield residual variability amplitudes of 1.95%, 1.87%, and 1.83% for the CO, CO + $PH_3$, and CO + $PH_3$ + $CO_2$ scenarios, respectively, all well above the estimated noise floor of 1.68%. The temperature-only model reduces the residual variability to 1.75%, demonstrating that thermal perturbations account for a large fraction of the observed modulation. The lowest residual variability is obtained for the combined temperature + VMR models, all of which reach 1.68%, effectively matching the noise floor. This indicates that once temperature variability is included, allowing CO to vary captures the remaining structured variability in the molecular-band regions, while additional $PH_3$ and $CO_2$ VMR variability produces no statistically measurable improvement.

Analysis of the retrieved temporal evolution shows that CO exhibits consistent and significant epoch-to-epoch variation across all models in which it is permitted to vary, indicating that CO abundance changes contribute measurably to the observed modulation. In contrast, $PH_3$ and $CO_2$ variations are smaller in amplitude across model configurations.

The primary driver of variability, however, appears to be temperature perturbation. In the temperature-only variability model, the largest changes occur at the pressure knot near $P \approx 0.45$ bar, where the retrieved temperature variations can reach up to $\sim$100 K. However, when CO VMR variability is included simultaneously, the amplitude of the temperature perturbation at this pressure level is reduced to $\sim$40 K, indicating that part of the spectral variability otherwise attributed solely to temperature can also be explained by changes in CO VMR. At deeper pressure levels, the inferred temperature variations remain much smaller, typically only a few kelvin. This suggests that the NIRSpec time-series spectra are most sensitive to variability in the upper atmospheric layers, where a combination of thermal perturbations and molecular VMR changes modulates the emergent spectrum. These results suggest that dynamical processes such as atmospheric circulation or vertical mixing produce spatially heterogeneous thermal/compositional structures that rotate into and out of view.

Overall, our results provide a retrieval-based framework for interpreting atmospheric variability in ultracold brown dwarfs. We demonstrate the following:

1. Thermal structure variability is the primary driver of time-dependent spectral modulation in WISE 0855−07.
2. Molecular abundance variations from CO, $PH_3$, and $CO_2$ provide a secondary contribution to the observed variability.
3. A cloud-free atmosphere is sufficient to explain the observed variability in WISE 0855−07 within the pressures probed by the dataset.

These findings place WISE 0855−07 firmly in the regime of dynamically active, thermally heterogeneous atmospheres analogous to solar system giant planets, while extending retrieval methodology to the coldest brown dwarf yet studied with JWST. Future time-series mid-infrared observations with JWST/MIRI will be critical for probing higher atmospheric layers where water-ice clouds are expected to reside and for determining whether cloud opacity variations contribute at longer wavelengths.

Time-resolved retrieval analyses such as this provide a powerful pathway toward connecting atmospheric dynamics, chemistry, and vertical structure in cold, planet-like objects. As JWST continues to deliver high-S/N time-series spectroscopy, similar approaches can be applied to other *Y* dwarfs and directly imaged exoplanets, enabling comparative studies of variability across the lowest-temperature substellar atmospheres.

## Acknowledgments

This material is based on work supported by NASA under grant No. 80NSSC24K0958 for the NASA XRP program. Support for program #JWST-GO-02327, JWST-AR-01977.004, JWST-GO-02124.009-A, and JWST-AR-03245.004-A was provided by NASA through a grant from the Space Telescope Science Institute, which is operated by the Associations of Universities for Research in Astronomy, Incorporated, under NASA contract NAS5-26555. We would also like to acknowledge the Texas Advanced Computing Center (TACC) at The University of Texas at Austin for providing computational resources that have contributed to the research results reported within this paper. URL: http://www.tacc.utexas.edu. J.M.V. acknowledges from a Royal Society—Research Ireland University Research Fellowship (URF/1/221932, RF/ERE/221108) and the European Union through the Exo-PEA ERC project (grant No. 101164652). Views and opinions expressed are however those of the author(s) only and do not necessarily reflect those of the European Union or the European Research Council Executive Agency. Neither the European Union nor the granting authority can be held responsible for them. Caroline V. Morley acknowledges: WISE 0855 JWST program: JWST-GO-02327 NASA XRP 80NSSC24K0958. The JWST data presented in this article were obtained from the Mikulski Archive for Space Telescopes (MAST) at the Space Telescope Science Institute. The specific observations analyzed can be accessed via DOI: 10.17909/yw1b-3765.

*Software:* Corner (D. Foreman-Mackey 2016), Matplotlib (J. D. Hunter 2007), Numpy (C. R. Harris et al. 2020).

## Appendix

Figure A1 shows 1D and 2D histograms for posterior PDFs for each retrieved parameter from the fiducial retrieval for WISE 0855–07.

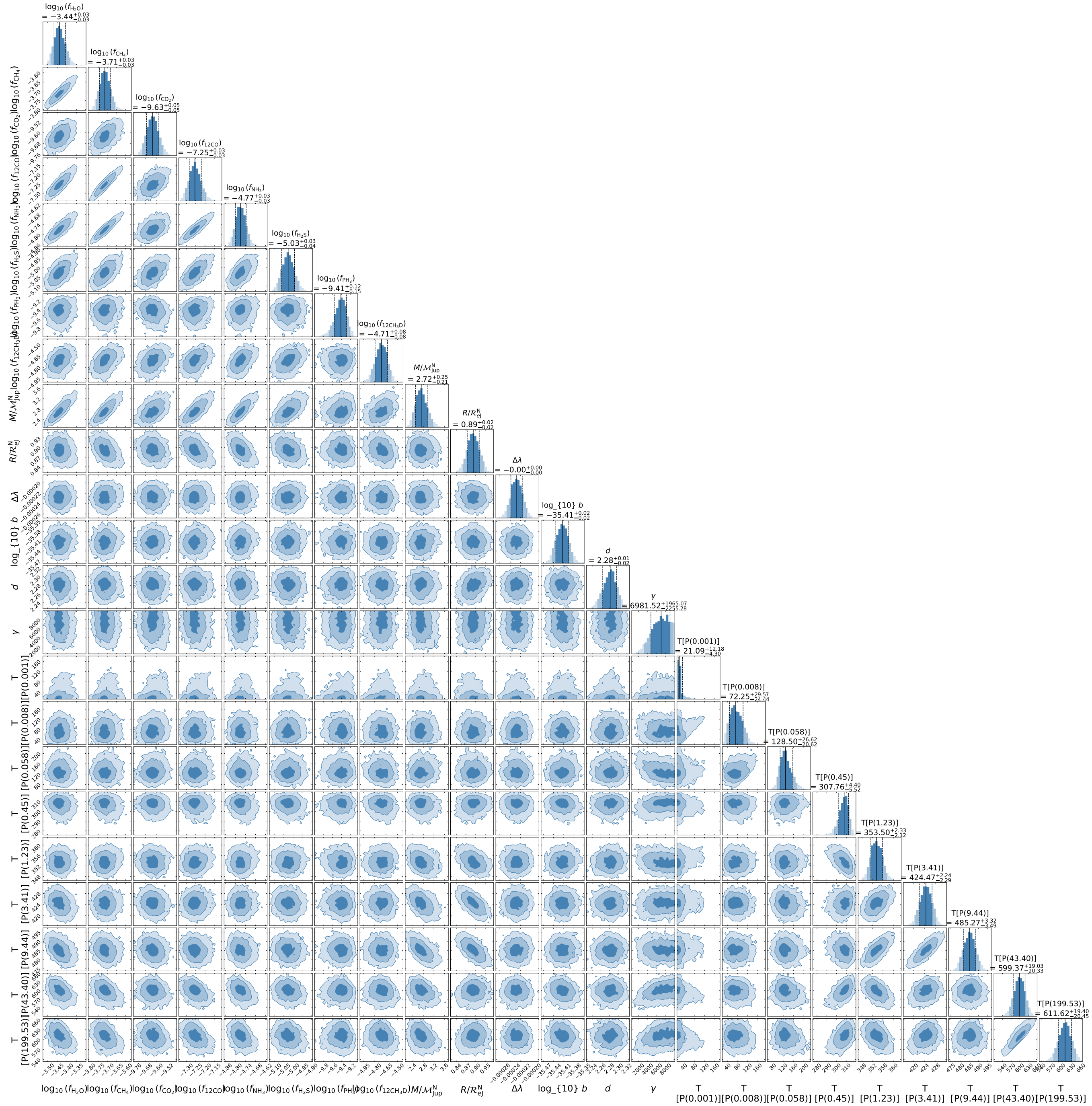

**Figure A1.** Marginalized posterior probability distributions for each parameter from the fiducial retrieval for WISE 0855–07. The first eight parameters represent the retrieved mixing ratios for $H_2O$, $CH_4$, $^{12}CO$, $CO_2$, $NH_3$, $H_2S$, $PH_3$, and $CH_3D$, followed by mass and radius. The parameters $\Delta\lambda$ and $\log b$ are nuisance parameters, $d$ is the distance to the object, and the last 10 parameters are used to calculate the thermal profile. The values above the 1D histograms represent the parametric median (50th percentile) values with the errors representing the $1\sigma$ central credible interval (16th and 84th percentile) values. The different shades in the 1D and 2D histograms represent the $1\sigma$, $2\sigma$, and $3\sigma$ central credible intervals, respectively, with the darkest shade corresponding to $1\sigma$.

## ORCID iDs

Harshil Kothari https://orcid.org/0009-0009-4489-0192
Caroline V. Morley https://orcid.org/0000-0002-4404-0456
Brittany E. Miles https://orcid.org/0000-0002-5500-4602
Melanie J. Rowland https://orcid.org/0000-0003-4225-6314
Natasha Batalha https://orcid.org/0000-0003-1240-6844
Michael C. Cushing https://orcid.org/0000-0001-7780-3352
Andrew J. Skemer https://orcid.org/0000-0001-6098-3924
James Mang https://orcid.org/0000-0001-5864-9599
Brianna Lacy https://orcid.org/0000-0002-9420-4455
Johanna M. Vos https://orcid.org/0000-0003-0489-1528
Channon Visscher https://orcid.org/0000-0001-6627-6067
Adam C. Schneider https://orcid.org/0000-0002-6294-5937
Genaro Suarez https://orcid.org/0000-0002-2011-4924
Mikayla J. Wilson https://orcid.org/0000-0003-3008-1975
Allison M. McCarthy https://orcid.org/0000-0003-2015-5029

## References

Ahrer, E.-M., Stevenson, K. B., Mansfield, M., et al. 2023, Natur, 614, 653
Apai, D., Radigan, J., Buenzli, E., et al. 2013, ApJ, 768, 121
Asplund, M., Grevesse, N., Sauval, A. J., & Scott, P. 2009, ARA&A, 47, 481
Azzam, A. A. A., Lodi, L., Yurchenko, S. N., & Tennyson, J. 2015, JQSRT, 161, 41
Beamín, J. C., Ivanov, V. D., Bayo, A., et al. 2014, A&A, 570, L8
Beer, R. 1975, ApJL, 200, L167
Biller, B. A., Vos, J. M., Zhou, Y., et al. 2024, MNRAS, 532, 2207
Buchner, J., Georgakakis, A., Nandra, K., et al. 2014, A&A, 564, A125
Buenzli, E., Apai, D., Radigan, J., Reid, I. N., & Flateau, D. 2014, ApJ, 782, 77
Burgasser, A. J., Gonzales, E. C., Beiler, S. A., et al. 2025, Sci, 390, 697
Burningham, B., Marley, M. S., Line, M. R., et al. 2017, MNRAS, 470, 1177
Burrows, A., Sudarsky, D., & Lunine, J. I. 2003, ApJ, 596, 587
Chamberlain, J. W., & Hunten, D. M. 1987, An Introduction to their Physics Andchemistry, Vol. 36 (Academic Press Inc.)
Faherty, J. K., Meisner, A. M., Burningham, B., et al. 2025, Natur, 645, 62
Faherty, J. K., Tinney, C. G., Skemer, A., & Monson, A. J. 2014, ApJL, 793, L16
Feroz, F., Hobson, M. P., & Bridges, M. 2009, MNRAS, 398, 1601
Foreman-Mackey, D. 2016, corner.py on GitHub, https://github.com/dfm/corner.py
Foreman-Mackey, D., Conley, A., Meierjurgen Farr, W., et al. 2013, emcee: The MCMC Hammer, Astrophysics Source Code Library, ascl:1303.002
Freedman, R. S., Lustig-Yaeger, J., Fortney, J. J., et al. 2014, ApJS, 214, 25
Hargreaves, R. J., Gordon, I. E., Rey, M., et al. 2020, ApJS, 247, 55
Harris, C. R., Millman, K. J., van der Walt, S. J., et al. 2020, Natur, 585, 357
Hogg, D. W., Bovy, J., & Lang, D. 2010, arXiv:1008.4686
Huang, X., Gamache, R. R., Freedman, R. S., et al. 2014, JQSRT, 147, 134
Hunter, J. D. 2007, CSE, 9, 90
Jakobsen, P., Ferruit, P., Alves de Oliveira, C., et al. 2022, A&A, 661, A80
Kirkpatrick, J. D., Gelino, C. R., Faherty, J. K., et al. 2021, ApJS, 253, 7
Kopytova, T. G., Crossfield, I. J. M., Deacon, N. R., et al. 2014, ApJ, 797, 3
Kothari, H., Cushing, M. C., Burningham, B., et al. 2024, ApJ, 971, 121
Kühnle, H., Patapis, P., Mollière, P., et al. 2025, A&A, 695, A224
Lacy, B., & Burrows, A. 2023, ApJ, 950, 8
Larson, H. P., Treffers, R. R., & Fink, U. 1977, ApJ, 211, 972
Lew, B. W. P., Roellig, T., Batalha, N. E., et al. 2024, AJ, 167, 237
Li, G., Gordon, I. E., Rothman, L. S., et al. 2015, ApJS, 216, 15
Line, M. R., Fortney, J. J., Marley, M. S., & Sorahana, S. 2014, ApJ, 793, 33
Line, M. R., Marley, M. S., Liu, M. C., et al. 2017, ApJ, 848, 83
Line, M. R., Teske, J., Burningham, B., Fortney, J. J., & Marley, M. S. 2015, ApJ, 807, 183
Lodders, K., & Fegley, B. 2006, Chemistry of Low Mass Substellar Objects (Springer), 1
Luhman, K. L. 2014, ApJL, 786, L18
Luhman, K. L., & Esplin, T. L. 2016, AJ, 152, 78
Luhman, K. L., Tremblin, P., Alves de Oliveira, C., et al. 2024, yCat, J/AJ/167/5
Mamajek, E. E., Prsa, A., Torres, G., et al. 2015, arXiv:1510.07674
McCarthy, A. M., Vos, J. M., Muirhead, P. S., et al. 2025, ApJL, 981, L22
Miles, B. E., Mang, J., & Morley, C. V. 2026, ApJ, submitted
Morley, C. V., Fortney, J. J., Marley, M. S., et al. 2012, ApJ, 756, 172
Morley, C. V., Marley, M. S., Fortney, J. J., & Lupu, R. 2014a, ApJL, 789, L14
Morley, C. V., Marley, M. S., Fortney, J. J., et al. 2014b, ApJ, 787, 78
Morley, C. V., Skemer, A. J., Allers, K. N., et al. 2018, ApJ, 858, 97
Nasedkin, E., Schrader, M., Vos, J. M., et al. 2025, A&A, 702, A1
Noll, K. S., Geballe, T. R., & Marley, M. S. 1997, ApJL, 489, L87
Polyansky, O. L., Kyuberis, A. A., Zobov, N. F., et al. 2018, MNRAS, 480, 2597
Radigan, J., Jayawardhana, R., Lafrenière, D., et al. 2012, ApJ, 750, 105
Richard, C., Gordon, I. E., Rothman, L. S., et al. 2012, JQSRT, 113, 1276
Rieke, G. H., Ressler, M. E., Morrison, J. E., et al. 2015, PASP, 127, 665
Rothman, L., Gordon, I., Barber, R., et al. 2010, JQSRT, 111, 2139
Rothman, L. S., Gordon, I. E., Barber, R. J., et al. 2010, JQSRT, 111, 2139
Rowland, M. J., Morley, C. V., Miles, B. E., et al. 2024, ApJL, 977, L49
Saumon, D., Geballe, T. R., Leggett, S. K., et al. 2000, ApJ, 541, 374
Saumon, D., Marley, M. S., Abel, M., Frommhold, L., & Freedman, R. S. 2012, ApJ, 750, 74
Schneider, A. C., Cushing, M. C., Kirkpatrick, J. D., & Gelino, C. R. 2016, ApJL, 823, L35
Sharp, C. M., & Burrows, A. 2007, ApJS, 168, 140
Skemer, A. J., Morley, C. V., Allers, K. N., et al. 2016, ApJL, 826, L17
Sousa-Silva, C., Al-Refaie, A. F., Tennyson, J., & Yurchenko, S. N. 2015, MNRAS, 446, 2337
Toon, O. B., McKay, C. P., Ackerman, T. P., & Santhanam, K. 1989, JGR, 94, 16287
Trotta, R. 2008, ConPh, 49, 71
Vos, J. M., Burningham, B., Faherty, J. K., et al. 2023, ApJ, 944, 138
Wang, F., Burningham, B., Littlefair, S., et al. 2026, MNRAS, 547, stag392
Wright, E. L., Eisenhardt, P. R. M., Mainzer, A. K., et al. 2010, AJ, 140, 1868
Yurchenko, S. N., & Tennyson, J. 2014, MNRAS, 440, 1649
Zalesky, J. A., Line, M. R., Schneider, A. C., & Patience, J. 2019, ApJ, 877, 24
Zapatero Osorio, M. R., Lodieu, N., Béjar, V. J. S., et al. 2016, A&A, 592, A80
Zhang, Z., Mollière, P., Fortney, J. J., & Marley, M. S. 2025, AJ, 170, 64